\documentclass[12pt, draftclsnofoot, onecolumn]{IEEEtran}
\usepackage{graphicx}
\usepackage{amsmath}
\usepackage{amssymb}
\usepackage{booktabs}
\usepackage{algorithm}
\usepackage{algorithmic}
\usepackage{subfig}
\usepackage{algorithm}
\usepackage{amsthm}
\usepackage{placeins}
\usepackage{amsfonts}

\usepackage[utf8]{inputenc}
\usepackage[english]{babel}
\usepackage{bm}
\newtheorem{theorem}{Theorem}[section]
\newtheorem{corollary}{Corollary}[theorem]
\newtheorem{lemma}[theorem]{Lemma}
\theoremstyle{definition}

\begin{document}

\title{Enhancing Security in Millimeter Wave SWIPT Networks}

\author{Rui~Zhu
\thanks{Rui Zhu is with the Department of Computer Science, Kettering University, 48504 USA (email: rzhu@kettering.edu)}
\thanks{}}


\maketitle

\begin{abstract}
Millimeter wave (mmWave) communication encounters a major issue of extremely high power consumption. To address this problem, the simultaneous wireless information and power transfer (SWIPT) could be a promising technology. The mmWave frequencies are more appropriate for the SWIPT comparing to current low-frequency wireless transmissions, since mmWave base stations (BSs) can pack with large antenna arrays to achieve significant array gains and high-speed short-distance transmissions. Unfortunately, the implementation of SWIPT in the wireless communication may lead to an expanded defencelessness against the eavesdropping due to high transmission power and data spillage. It is conventionally believed that narrow beam offers inherent information-theoretic security against the eavesdropping, because only the eavesdroppers, which rely on the line-of-sight path between the legitimate transmitter and receiver, can receive strong enough signals. However, some mmWave experiments have shown that even by using highly directional mmWaves, the reflection signals caused by objects in the environment can be beneficial to the eavesdroppers. This paper studies the security performance in general mmWave SWIPT networks, and investigates the probability of successful eavesdropping under different attack models. Analytical expressions of eavesdropping success probability (ESP) of both independent and colluding eavesdroppers are derived by incorporating the random reflection paths in the environment. Theoretical analysis and simulation results reveal the effects of some key parameters on the ESP,  such as the time switching strategy in SWIPT, densities of mmWave BSs, and carriers frequencies, etc. Based on the numerical and simulation results, some design suggestions of mmWave SWIPT are provided to defend against eavesdropping attacks and achieve secure communication in practice.
\end{abstract}


\section{Introduction}
Millimeter wave (mmWave) communication is a key technique for the next generation (5G) wireless networks, which can leverage large spectrum resources at higher frequencies to achieve much higher data rates. But at mmWave frequencies, the signals exhibit high propagation losses. To overcome this problem, mmWave communication systems adopt spatial beamforming using large antenna array, and increase the transmission power at the mmWave transmitter. Since the size of an antenna is determined by the wavelength, smaller wavelength means smaller size of each antenna. Therefore, large-scale antenna arrays can be integrated into a mmWave transceiver, which can achieve significant array gains. However, the implementation of large antenna arrays and the increase of power transmission lead to higher power consumption, which will be a big concern for power-limited wireless devices in future applications such as the Internet of Things (IoT) \cite{6740844}. 

To address the extremely high power consumption issue in the millimeter wave (mmWave) communication, the simultaneous wireless information and power transfer (SWIPT) could be a promising technology. SWIPT is a recently developed technique, that enables the simultaneous transfer of information and power wirelessly. The receiver can harvest the energy from the received radio frequency (RF) signals to charge its battery or to transmitter signals. The mmWave frequencies are more appropriate for the SWIPT when compared with current low-frequency wireless transmissions, since mmWave base stations (BSs) can pack with large antenna arrays to achieve significant array gains and high-speed short-distance transmissions. 

Besides the concern of power consumption, secure communication is also a key factor in the quality of experience (QoE) in 5G. Unfortunately, in the security perspective, the implementation of SWIPT in the wireless communication may lead to an expanded defencelessness against the eavesdropping due to high transmission power and data spillage \cite{8214104}. Since in order to expedite the energy harvest process, the transmitter usually emits a highly boosted signal. Conventionally, it is believed that narrow beam offers inherent information-theoretic security against the eavesdropping, because only the eavesdroppers that rely on the line-of-sight path between the legitimate transmitter and receiver can receive strong enough signals. However, some mmWave experiments have shown that even by using highly directional mmWaves, the reflection signals caused by objects in the environment will be beneficial to the eavesdroppers \cite{7346844}.

To the best of our knowledge, the security performance of mmWave simultaneous wireless
information and power transfer (SWIPT) ultra-dense networks is still not well investigated, which motivates this research. The analytical study of the security performance in mmWave SWIPT networks is challenging, since 1) channel behaviours at different mmWave frequencies are still under scrutiny, and there is no any standard channel model; 2) in contrast with current low-frequencies, the signal propagation at mmWave frequencies is sensitive to blockage and reflection by the objects in environment, which introduce more randomness in the analysis of security; and 3) compared to the conventional mmWave ad hoc networks, by incorporating the power transfer, the signal that is for energy harvest should be counted as additional noise to the eavesdroppers.

In this paper, we adopt time switching policy receiver architectures at the users to harvest energy from the received
RF signals and process the information. The SWIPT mmWave ultra-dense networks are modeled under a stochastic geometry framework. In this model, we derive the eavesdropping success probability (ESP) of eavesdroppers under independent and colluding eavesdropping attack strategies.

\subsection{Related Work}
\subsubsection{Secure communication in mmWave networks}
Information-theoretic security in mmWave point-to-point, ad hoc networks and vehicular
communication systems have been investigated in \cite{7862142}, \cite{7880674}, and \cite{7876781}, respectively. In \cite{7862142}, the secure transmissions under slow fading channels in a mmWave system were investigate. The authors analyzed the secrecy properties of the maximum ratio transmitting beamforming for the mmWave system, and proposed two secure transmission schemes, namely AN beamforming and partial MRT. But they did not consider the geometry in the mmWave dense networks. In \cite{7880674}, the authors studied the physical layer security in the mmWave ad hoc networks. In this work, the mmWave ad hoc networks were modeled with the help of stochastic geometry. The effect of blockage was also incorporated. The average achievable secrecy rate was derived to quantify the impacts of key system parameters on the secrecy performance. In \cite{7876781}, two physical layer security techniques for vehicular mmWave communication systems were proposed. They used multi-antenna to transmit information symbols and artificial noise in target receiver and non-receiver directions, respectively.

The secrecy performance of mmWave overlaid micro-wave networks and mmWave cellular network have been analyzed in \cite{7464352} and \cite{7505974}, respectively. To date, there exists several works in the literature that studied the eavesdropping attack in mmWave networks. The authors in \cite{7178504} developed an attack on the antenna subset modulation technique. In \cite{7346844}, authors studied the impact of reflections on the physical layer security of mmWave systems. They experimentally demonstrated that the eavesdropper has successfully eavesdropped, although the legitimate transmitter used highly directional signal beams in the presence of reflection paths. The work in \cite{8723482} discussed eavesdropper attack strategies for 802.11ad mmWave systems and provided the first analytical model to characterize the success possibility of eavesdropping in both opportunistic stationary attacks and active nomadic attacks. The success probability of eavesdropping considering the ambient reflectors in the environment and beam misalignment errors were derived. Compared with \cite{8723482}, our work studies the eavesdropping success probability in a more general model, incorporating the SWIPT and the interference from other mmWave BSs. For the SWIPT networks, the signal that is used for the energy harvest is additional noise to jam the eavesdroppers. In addition, the interference from other random-distributed mmWave BSs can not be ignored, which makes the analysis of eavesdropping success probability more complicated. Moreover, we propose two attack strategies, named independent eavesdropping and colluding eavesdropping. 

\subsubsection{SWIPT in mmWave networks}
The wireless power transfer in mmWave networks has attracted an increasing amounts of research in recent year, since the mmWave networks are suitable for the wireless power transfer \cite{8387202}\nocite{7491259}\nocite{Wang_2017}-\cite{7997004}. In \cite{8387202}, the authors investigated the benefits of heterogeneous ultra-dense network architecture from the perspective of wireless energy harvesting, and provided an enthusiastic outlook toward application of WPT in heterogeneous ultra-dense networks. In \cite{7491259}, the authors studied the approximate energy coverage probability, energy-information coverage probability, and the average harvested power of mmWave SWIPT networks under power splitting policy. In \cite{Wang_2017}, the required small cell density to achieve an expected level of harvested energy was obtained. In \cite{7997004}, the energy coverage and mode selection mechanism in sub-6 GHz were studied. These studies were focused on the performance of energy harvest, and no security issues have been considered in them. 

\subsubsection{Information-theoretic security in SWIPT networks}
The problems of information-theoretic security and energy efficiency for SWIPT two-hop wireless networks have been investigated in our previous work \cite{8500123}. In \cite{8500123}, we studied the problem of power allocation to maximize the transmission energy efficiency (TEE), while satisfying the security requirements in a general wireless relay network. In this work, we study a different configuration of the problem by focusing on the secure communication in a one-hop mmWave network, such as a mmWave cellular network. Due to the characteristics of mmWave, e.g. blockage, it is more challenging to study the security in mmWave SWIPT networks. In \cite{8566013}, Sun et al. considered the secure communications in downlink SWIPT mmWave ultra-dense networks. The analytical expressions of the energy-information coverage probability were derived. Then, the closed-form expressions of the secrecy probability in the presence of multiple independent or colluding eavesdroppers, and the effective secrecy throughput (EST), were derived. However, the effect of reflectors in the real environment and active eavesdropping attacks were not considered. 

\subsection{Contributions and Organization}
This paper studies information-theoretic security in ultra-dense mmWave information and power transfer networks. Our analysis accounts for the new features of mmWave channel and the effects of line-of-sight (LOS) and reflection paths in signal propagation. The main contributions of this work are summarized as follows.

\begin{itemize}
  \item We model the mmWave simultaneous information and power transfer (SWIPT) networks under stochastic geometry framework, to characterize the random spatial locations of mmWave, mobile users, eavesdroppers and obstacles in the real environments. The effect of reflection and blockage are incorporated such that the signal propagation links are either LOS or reflection.
  \item In our analytical model, we consider the random antenna beam directions on both mmWave BSs and eavesdroppers, and the errors introduced by the beam direction misalignment between BSs and eavesdroppers.
  \item Compared with the conventional mmWave ad hoc networks, the model of mmWave simultaneous information and power transfer (SWIPT) networks in this study is a more general model, by incorporating the power transfer. The RF signal that is used for power transfer can be considered as an additional noise to jam eavesdroppers. The conventional mmWave ad hoc network is an special case of our model, where the time switching ratio is zero.
  \item Our analysis models two types of eavesdropping attack strategies to the mmWave SWIPT networks, namely independent eavesdropping and colluding eavesdropping attacks. We derive the analytical expressions for eavesdropping success probability (ESP) of eavesdroppers under independent and colluding eavesdropping attacks, considering both the LOS and reflection paths of signal propagation. 
  \item To get insights, we statistically examine the security performance trends in terms of key parameters such as the time switching strategy in SWIPT, densities of mmWave BSs, and carriers frequencies, etc. Our numerical and simulation results reveal the effects of some key parameters on the security performance, for example, the larger time switching ratio $\eta$ leads to lower probability of success eavesdropping attacks.
\end{itemize}

The reminder of this paper is organized as follows. In Section \ref{mm_model}, we introduce the system model and performance metric we study in this paper. Section \ref{mm_esp} presents two eavesdropping attack strategies for the mmWave SWIPT networks and the discussion of eavesdropping success probability for Eves under these strategies. In Section \ref{mm_results}, we provide the analytical validation of our proposed attack models by numerical results, and provide simulation results to support the analytical models. We concluded this paper in Section \ref{mm_conclusion}.

\section{System Model}
\label{mm_model}
\subsection{Network and SWIPT Model}

\begin{figure}[htbp]
	\centering
	\includegraphics[scale=0.3]{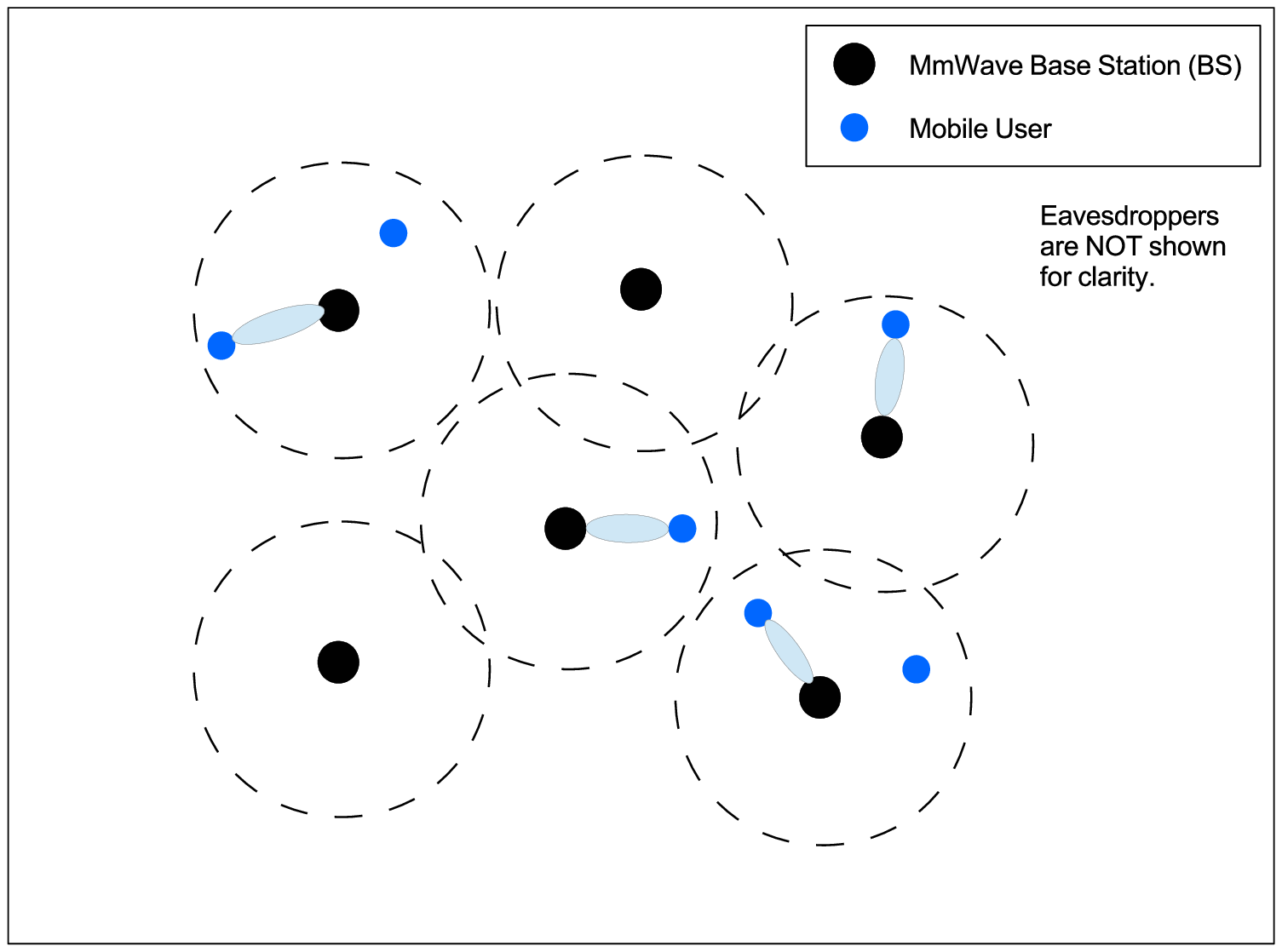}
	\caption{Downlink multiuser multicell interference scenario in a ultra-dense mmWave network. For clarity, eavesdroppers, the intercell and intracell interference are not shown in this figure.}\label{system_model}
\end{figure}

In this paper, a mmWave ultra-dense wireless SWIPT network consisting of mmWave BSs, a population of wireless-powered mobile devices (users), and malicious eavesdroppers that operate in the mmWave band is considered, as shown in Figure~\ref{system_model}. In this network, authorized mobile users extract energy and confidential information from spatially distributed BSs in the presence of multiple malicious eavesdroppers. And only the time switching (TS) SWIPT policy is considered in this work \cite{8214104}. For TS policy, authorized mobile users use a fraction of the block time to harvest the energy and the remaining time for information processing. Specifically, we consider the users harvest the energy in the first $\eta T$ duration, and process information in the remaining $(1-\eta)T$ duration, where $T$ is the coherent time, $\eta \in (0,1)$ is the fraction of the block time for users' information processing, which is called time switching ratio.

The mmWave BSs, mobile users and eavesdroppers are randomly distributed according to homogeneous Poisson point processes (HPPPs) $\phi_{b}$, $\phi_{u}$, and $\phi_{e}$ with density $\lambda_{b}$, $\lambda_{u}$ and $\lambda_{e}$, respectively. Due to the high penetration losses of mmWave signals for common building materials \cite{6932503}, we can assume that the building blockages are impenetrable.

We split the user population into two types of users, namely connected and nonconnected users. A connected user is assumed to be tagged with a BS that maximizes the average received power at that user. Moreover, for the connected user, we assume perfect beam alignment between a BS and this user, i.e., the BS and user point their beams so as to have the maximum gain. Furthermore, we assume that a BS serves only one connected user at a given time. Therefore, due to limited resources, the mmWave network may serve only a fraction of the user population as connected users, leaving the rest in the nonconnected mode. In this work, only the connected users are considered. Moreover, we consider that mobile users always connect the BS that offers minimum path loss to it as serving BS \cite{6932503}.

\subsection{Obstacles and Reflectors Assumptions}
In this work, we uniformly model the obstacles and reflectors as rectangular objects. Let $R$, $l$, $w$, and $\theta_{R}$ denote each object's (obstacle or reflector) center location, length, width, and orientation, respectively, where $\theta$ is the anti-clockwise angle between the x-axis and the length of the object. The centers $R$ of objects follow a homogeneous Poisson Point Process (PPP) $\Phi$ with density $\lambda_{0}$. The lengths and widths of objects are distributed with PDF $f_{L}(l)$ and $f_{W}(w)$, respectively.  

Only the reflection paths that follow specular reflection law are considered in this work. The diffusion and diffraction paths are neglected due to the fact that these propagation mechanisms do not contribute significantly to the received signal power at mmWave frequencies \cite{5783993}. Let $\rho$ denote the reflection coefficient of reflectors, $P_{i}$ and $P_{r}$ denote the powers of the incident and reflection signals, respectively. The reflection signal power is expressed as
\begin{equation}
P_{r}=\rho \cdot P_{i}.
\end{equation}

\subsection{Channel Model}
We now describe the channel model for an arbitrary user. Empirical evidence suggests that mmWave frequencies exhibit different propagation characteristics for the LOS/NLOS links \cite{6932503}. While the LOS mmWave signals propagate as if in free space, the NLOS mmWave signals typically exhibit a higher path loss exponent (and additional shadowing) \cite{6932503}. We let $\alpha_{L}$ and $\alpha_{N}$ be the path loss exponents for the LOS or NLOS links respectively. Given a distance $|X|$, the path loss function is defined as follows:

\begin{equation}
L(|X|)=b(max(d,|X|))^{-\alpha}
\end{equation}
where $\alpha=\alpha_{L}$ for LOS link and $\alpha=\alpha_{N}$ for NLOS link, $d$ is a reference distance \cite{1580787}; $b$ is a frequency independent constant parameter of path loss, commonly set as $(\frac{c}{4\pi f_{c}})^{2}$ with carrier frequency $f_{c}$, and $\beta=\beta_{L}$ for LOS link and $\beta=\beta_{N}$ for NLOS link.

\subsection{Antenna Model}
\begin{figure}[htbp]
	\centering
	\includegraphics[scale=0.15]{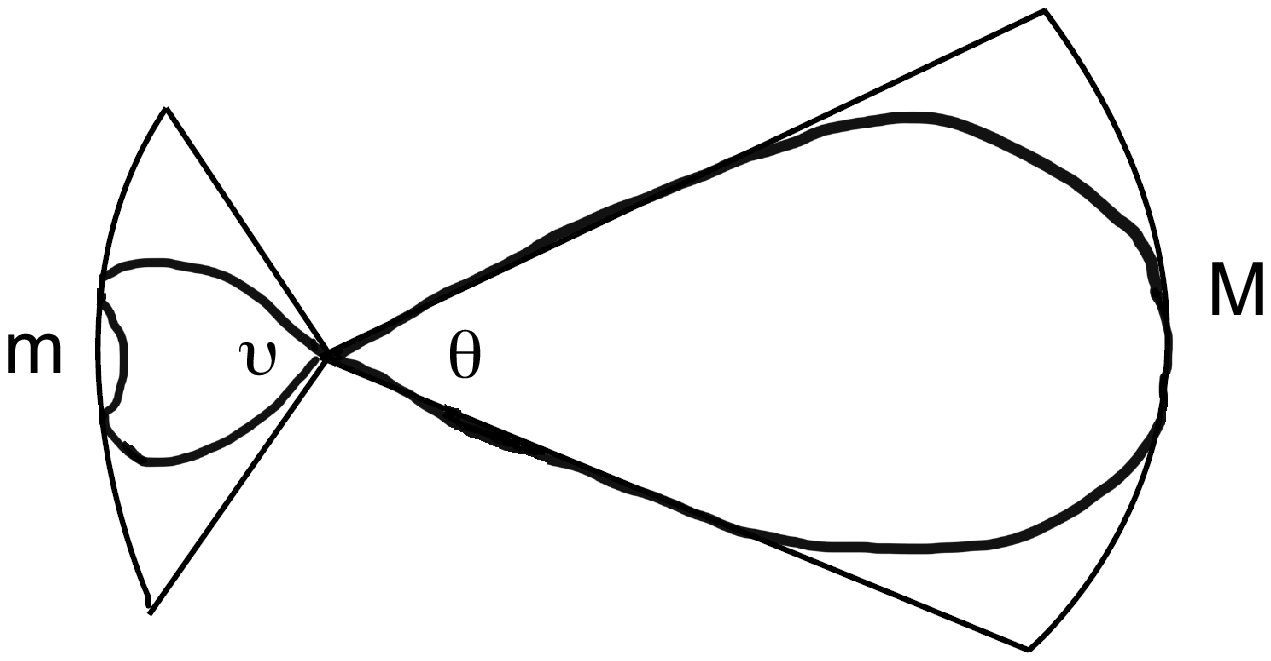}
	\caption{Sector antenna model. $M, m, \theta, \upsilon$ denote the main lobe, side lobe, the half power beamwidth for the main lobe, and the half power beamwidth for the side lobe, respectively.}\label{antenna}
\end{figure}

To compensate for higher propagation losses, mmWave BSs use large directional antennas arrays \cite{7880674}. In this study, we consider the directional beamforming and use an ideal sector model to analyze the beam pattern, as shown in Figure~\ref{antenna}.

Let $G_{ideal}(\omega)$ be an ideal sector antenna, the ideal sector antenna model is given by

\begin{equation}
\label{antenna_model}
G_{ideal}(\omega)=
\left\{
\begin{array}{lr}
G_{m}, & \text{if}\ \omega \le \frac{\theta}{2}, \\
G_{s}, & \text{else.}
\end{array}
\right.
\end{equation}
where $\theta$ is the mainlobe beamwidth; $G_{m}$ and $G_{s}$ are the mainlobe gain, and the sidelobe gain over the mainlobe beamwidth, and the side beamwidth, respectively. 

\subsection{Antenna Gain Distribution with Beam Misalignment}
In this work, we assume that a legitimate mmWave BS and its connected user has perfect beam alignment. However, the beam misalignment between the BS and an eavesdropper is likely to occur, as shown in Figure~\ref{misalignment}. Let $\theta_{\epsilon}$ denote the antenna pattern misalignment. For simplicity, we only consider $\theta_{\epsilon}$ as a positive value. 

\begin{lemma}
\label{lemma1}
The antenna gain PDF $f_{G_{e}}(x)$ that is obtained by an arbitrary eavesdropper for $x \in [G_{m}, G_{s}]$ is given by
\begin{equation}
\label{pdf_gain}
f_{G_{e}}(x)=F_{\theta_{\epsilon}}(\frac{\theta}{2})\delta(x-G_{m})+(1-F_{\theta_{\epsilon}}(\frac{\theta}{2}))\delta(x-G_{m})
\end{equation}
where $F_{\theta_{\epsilon}}(x)$, $x \le \frac{\theta}{2}$ is the CDF of beam misalignment $\theta_{\epsilon}$.
\end{lemma}

\begin{proof}
\begin{figure}[htbp]
	\centering
	\includegraphics[scale=0.3]{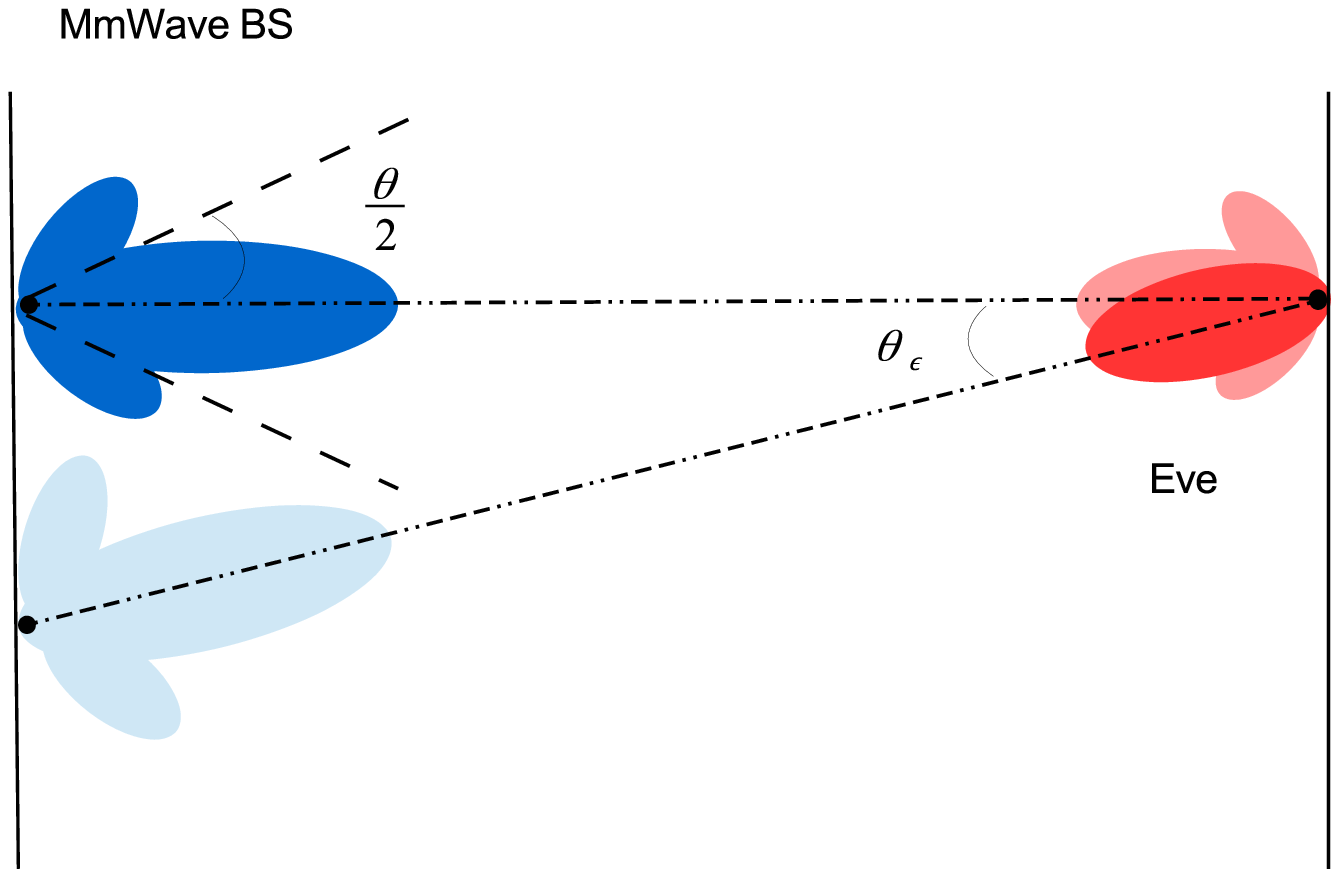}
	\caption{Antenna beam pattern misalignment between a mmWave BS and an eavesdropper. $\theta$ denotes the mainlobe beamwidth, and $\theta_{\epsilon}$ denotes the misalignment.}\label{misalignment}
\end{figure}

By following the beam misalignment error model proposed in \cite{8723482} \cite{article} \cite{6835179}, the beam alignment $\theta_{\epsilon}$ is a random variable, which is bounded within the range of mainlobe beamwidth $\theta$, i.e., $0 \le \theta_{\epsilon} \le \frac{\theta}{2}$. And $\theta_{\epsilon}$ follows truncated normal distribution, and the PDF of $\theta_{\epsilon}$ is given by \cite{8723482} as follows:

\begin{equation}
f_{\theta_{\epsilon}}(x) = \frac{e^{\frac{-x^{2}}{2\sigma_{\theta_{\epsilon}}^{2}}}}{\sigma_{\theta_{\epsilon}}\sqrt{2\pi} \cdot erf(\frac{\theta}{2\sqrt{2}\sigma_{\theta_{\epsilon}}})}
\end{equation}
where $\theta_{\epsilon}^{2}$ is the variance of the angle error.

The CDF of $\theta_{\epsilon}$ is given by
\begin{equation}
\begin{aligned}
F_{\theta_{\epsilon}}(x) &= \int_{0}^{\frac{\theta}{2}} f_{\theta_{\epsilon}}(x) dx\\
&=\int_{0}^{\frac{\theta}{2}}\frac{e^{\frac{-x^{2}}{2\sigma_{\theta_{\epsilon}}^{2}}}}{\sigma_{\theta_{\epsilon}}\sqrt{2\pi} \cdot erf(\frac{\theta}{2\sqrt{2}\sigma_{\theta_{\epsilon}}})}dx
\end{aligned}
\end{equation}

Since (\ref{antenna_model}) produces either gains $G_{m}$ or $G_{s}$ over all possible input angles, the resulting gain distributions are discrete. Therefore, the ideal sector antenna model can be rewritten as 
\begin{equation}
\label{antenna_model_re}
G_{ideal}(\omega)=
\left\{
\begin{array}{lr}
G_{m}, & Pr=F_{\theta_{\epsilon}}(\frac{\theta}{2}), \\
G_{s}, & Pr=1-F_{\theta_{\epsilon}}(\frac{\theta}{2}).
\end{array}
\right.
\end{equation}

From (\ref{antenna_model_re}) and apply the dirac delta function, we can derive the PDF $f_{G_{e}}(x)$ as 
\begin{equation}
f_{G_{e}}(x)=F_{\theta_{\epsilon}}(\frac{\theta}{2})\delta(x-G_{m})+(1-F_{\theta_{\epsilon}}(\frac{\theta}{2}))\delta(x-G_{m})
\end{equation}
where $F_{\theta_{\epsilon}}(x)$, $x \le \frac{\theta}{2}$ is the CDF of beam misalignment $\theta_{\epsilon}$.
\end{proof}

\begin{corollary}
\label{corollary1}
By Lemma \ref{lemma1}, the expectation of antenna gain that is obtained by an arbitrary eavesdropper is given by
\begin{equation}
\begin{aligned}
\mathbb{E}[G_{e}] &=\int_{-\infty}^{\infty}xf_{G_{e}}(x)dx \\
&=F_{\theta_{\epsilon}}(\frac{\theta}{2})G_{m}+(1-F_{\theta_{\epsilon}}(\frac{\theta}{2}))G_{s}.
\end{aligned}
\end{equation}
\end{corollary}

\subsection{Stochastic Geometry Framework}
\subsubsection{LOS Link Distribution}
In this work, the obstacles and reflectors in the environment are modeled as a PPP. The total number of obstacles and reflectors $N$ between BS and an arbitrary eavesdropper separated by distance $d$ is a Poisson random variable with the expectation given by \cite{6840343}
\begin{equation}
\mathbb{E}[N]=\beta_{0}d+p
\end{equation}
where $\beta_{0}=\frac{2\lambda \mathbb{E}[L] + \mathbb{E}[W]}{\pi}$, and $p=\lambda \mathbb{E}[L] \mathbb{E}[W]$.

\begin{figure}[htbp]
	\centering
	\includegraphics[scale=0.3]{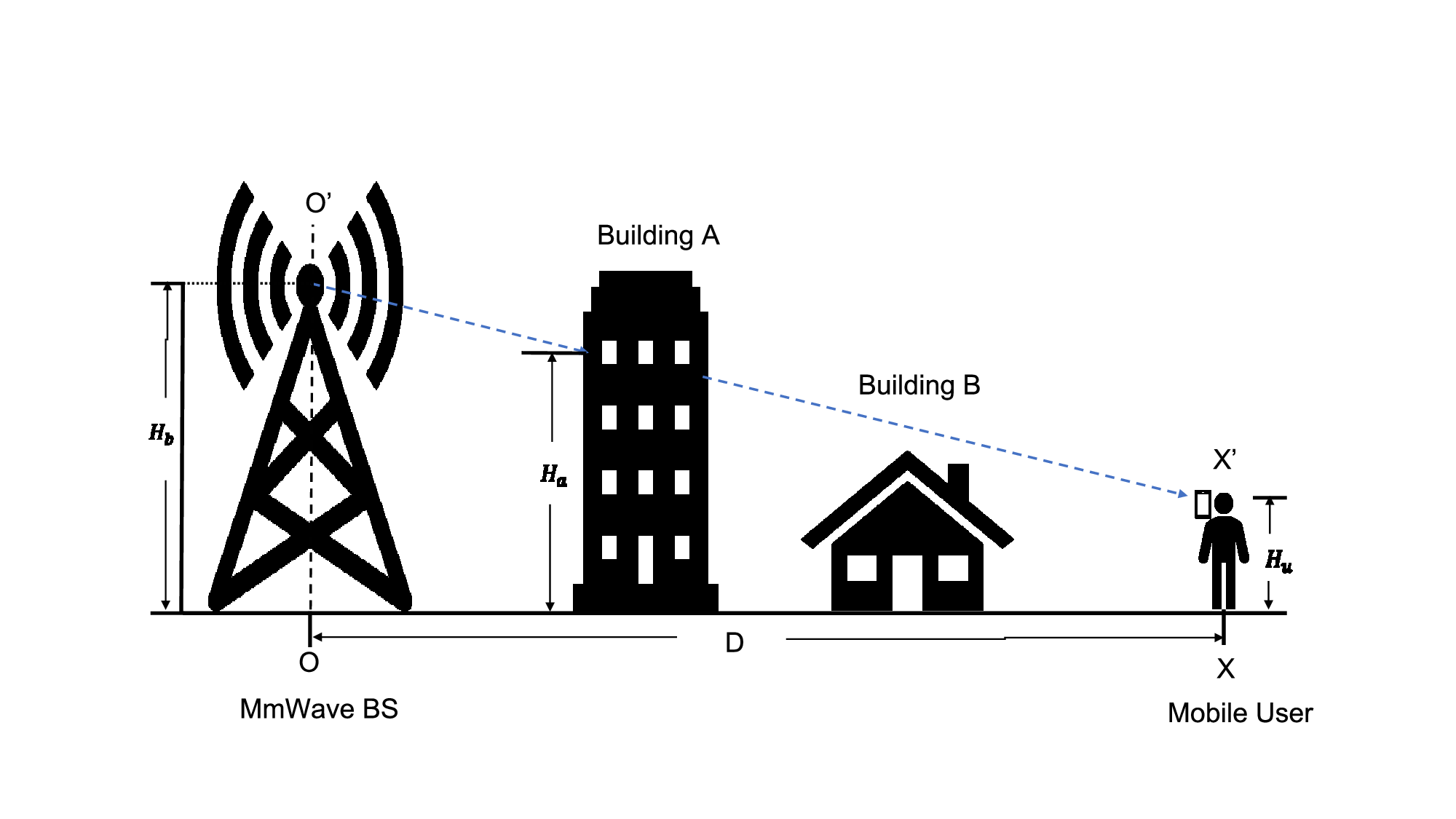}
	\caption{One realization of blockage in the mmWave communication in the real environment, where the mmWave BS located at $O$ has a height $H_{b}$, while the mobile user has a height $H_{u}$. \cite{6840343} }\label{blockage_los}
\end{figure}

Considering the height of obstacles, we should note that not all obstacles between BS and an arbitrary eavesdropper can block the actual signal propagation in $\mathbb{R}^{3}$, such as the building B in Figure~\ref{blockage_los}. In this figure, the mmWave BS located at $O$ has a height $H_{b}$, while the mobile user has a height $H_{u}$. Only a building intersecting $OX$ will block the signal propagation over path $O'X'$ if and only if its height larger than $H_{a}$. In this work, the height of obstacles $H_{i} \in [H_{min}, H_{max}]$ is a random variable with the PDF $f_{H}(h)$.

Therefore, incorporating the height of blockages, \cite{6840343} introduces a constant scaling factor $k$, and derives the probability that there exists a LOS link between the BS and an arbitrary Eve  with distance $d$, i.e., the number of obstacles and reflectors between them is $0$ as follows

\begin{equation}
\label{eq_plos}
P_{Los}=e^{-k(\beta_{0}d+p)},
\end{equation}
where $k=1-\int_{0}^{1}\int_{H_{min}}^{sH_{u}+(1-s)H_{b}}f_{H}(h)dhds$.

\subsubsection{Reflection Link Distribution}
Let $l$ and $w$ denote the length and width of an arbitrary obstacle or reflector in the environment. The reflector orientation $\theta_{R}$ is a random variable. 

The probability that there exists a first-order reflection path between the BS and an arbitrary Eve, with propagation delay $\tau$ and reflection point located at $R$, that is not blocked by other obstacles is given by \cite{Muhammad2017AnalyticalMF}

\begin{equation}
\label{eq_pref}
\begin{aligned}
P_{Ref} &=exp(\lambda_{0}(\mathbb{E}[l]\sqrt{c^{2}\tau^{2}-D^{2}cos^{2}\theta}\\
&+\mathbb{E}[w]D|cos\theta|+\mathbb{E}[l]\mathbb{E}[w]-\frac{\mathbb{E}[l](c\tau-D)}{4} \\
&-\frac{\mathbb{E}[l]^{2}\sqrt{c^{2}\tau^{2}-D^{2}}}{8D})). 
\end{aligned}
\end{equation}

\begin{figure}[htbp]
	\centering
	\includegraphics[scale=0.3]{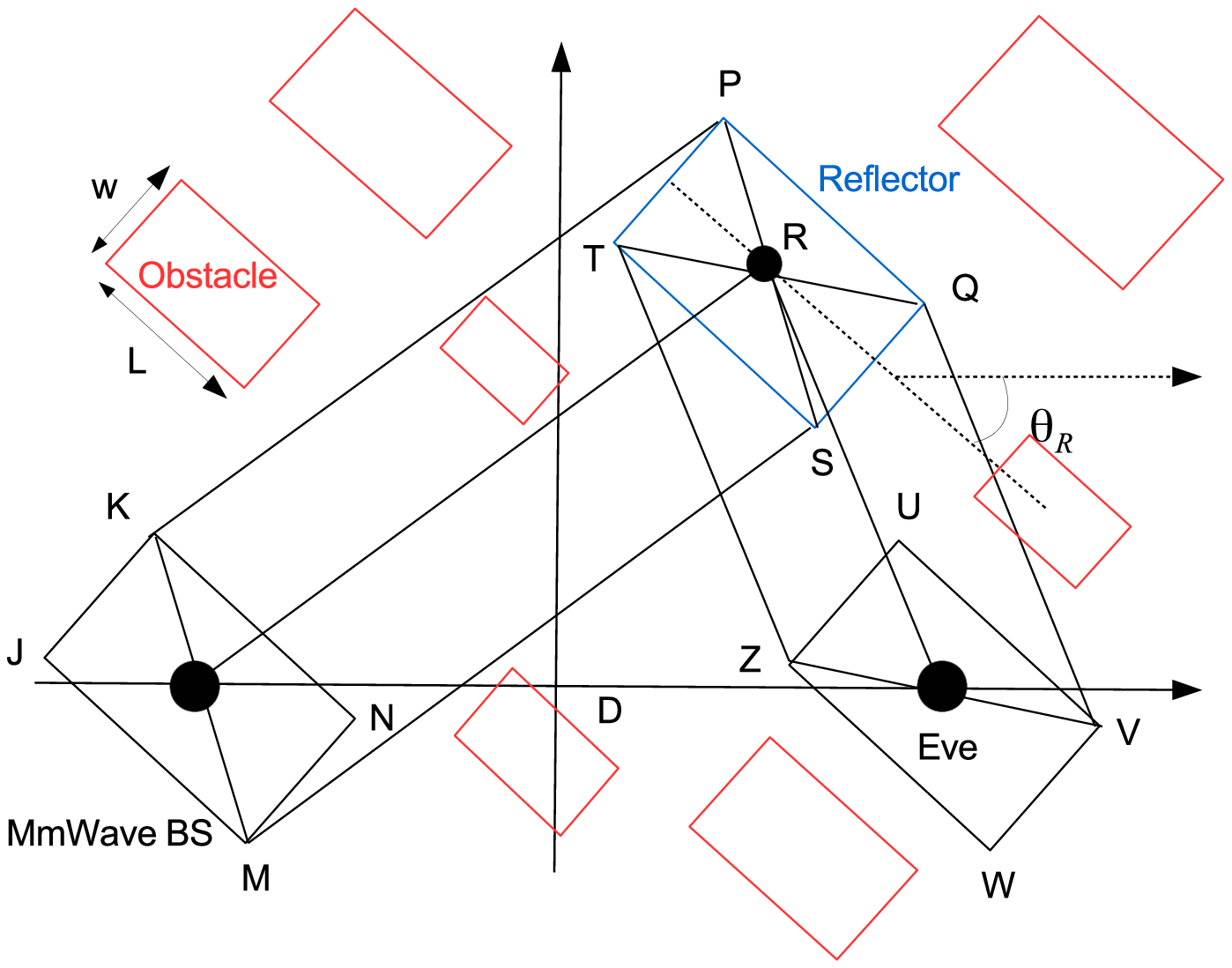}
	\caption{Geometry of the reflection path with blockage. \cite{Muhammad2017AnalyticalMF} }\label{blockage_ref}
\end{figure}

\subsubsection{Distance Distribution}
Since the mmWave BSs and eavesdroppers are randomly distributed according to homogeneous Poisson point processes (HPPPs) $\phi_{b}$ and $\phi_{e}$, the distance between a typical BS and an eavesdropper $d$ is a random variable. The cumulative density function (CDF) of the random variable $d$ is given by the following lemma~\cite{8723482}:

\begin{lemma}
\label{lemma2}
The CDF $F_{d}(x)$ of the random variable $x$ is given by

\begin{equation}
F_{d}(x)=1-e^{-2\pi \lambda_{b} \frac{e^{-p}}{\beta_{0}^{2}} (1-\beta_{0}xe^{-\beta_{0}x}-e^{-\beta_{0}x})}
\end{equation}
where $\beta_{0}=\frac{2\lambda_{0}(\mathbb{E}[L] + \mathbb{E}[E])}{\pi}$, and $p=\lambda_{0} \mathbb{E}[L] \mathbb{E}[E]$, and $\lambda_{b}$ denotes the intensity of the BSs.
\end{lemma}
\begin{proof}
The proof can be found in \cite{6840343}.
\end{proof}

Consider a PPP in 2D space $\mathbb{R}^{2}$ with density $\lambda$ per unit area. Let $B_{o,r}$ denote a disk centered at origin with radius r. Let $n$ be the number of points inside the disk $B$. Given $n \ge 1$, let $\{x_{1}, x_{2},...,x_{n}\}$ be the set of distances of points inside the disk respect to the origin. The distribution of distance $x$ is given by the following lemma.

\begin{lemma}
\label{lemma3}
Given $n$, the PDF $f_{r}(x)$ of the random variable $x$ is given by

\begin{equation}
f_{r}(x)=\frac{2x}{r^{2}}
\end{equation}
\end{lemma}

\begin{proof}
Given $n$, the $n$ points are uniformly distributed throughout $B$. Since a circle's circumference is proportional to its radius, the density $f_{r}$ of $x_{i}$ for all $i$ is also proportional to its radius. That is $f_{r}(x)=ax$, for some constant $a$. Then solve for $a$ as follows:
\begin{equation}
\begin{aligned}
1 &= \int_{0}^{r}axdx =[\frac{ax^{2}}{2}]^{R}_{0} =\frac{ar^{2}}{2}
\end{aligned}
\end{equation}
So $a=\frac{2}{r^{2}}$, and $f_{r}(x)=\frac{2x}{r^{2}}$.
\end{proof}

\subsection{Attack Models}
We consider the case of passive eavesdropping without any active attacks to deteriorate the information transmission. The locations of eavesdroppers are modeled following an independent homogeneous PPP $\phi_{e}$ with $\lambda_{e}$.

In this work, two attack models are considered: (1) the independent eavesdroppers that can try to eavesdrop from
its location independently, and (2)
the colluding eavesdroppers that can jointly process their received signals at a central data processing unit. In the second attack model, we consider the eavesdroppers adopt the maximal-ratio combining (MRC) scheme to get the most confidential information and the best possible performance. The worst case of eavesdropping attacks is considered in both of these two scenarios that all received RF signals by the eavesdroppers are used for information processing.

\subsection{Performance Metric}
In this work, we study the \textbf{Eavesdropping Success Probability (ESP)} as the performance metric. The ESP is used to characterize
the secure transmission between the typical user and
its serving BS, that is no information leakage to eavesdroppers. And it is defined as the probability that the signal-to-interference-plus-noise ratio (SINR) of an eavesdropper Eve is greater than a certain threshold that Eve can successfully overhear the signal from a typical mmWave BS. Formally,  we define $C_{Eve}: SINR_{Eve}>\beta$ as the event when the SINR of Eve is greater than the threshold $\beta$. Accordingly, the probability that eavesdroppers will successfully overhear the signal is given by
\begin{equation}
P(C_{es}) = P(SINR_{Eve}>\beta)
\end{equation}

\section{Eavesdropping Success Probability}
\label{mm_esp}
In this study, we adopt the Time switching (TS) strategy for the information and power transfer between the mmWave BSs and mobile users: the BSs send energy for the first $\eta T$ duration and
send confidential information for the remaining $(1-\eta) T$ duration,
where $T$ is the coherent time, $0 <\eta< 1$ is the fraction of
time for the BSs' power transfer, and $1-\eta$ is the fraction of time for users’ confidential information receiving. 

\subsection{Independent Eavesdroppers}
In the independent eavesdroppers model, the eavesdroppers independently overhear BS transmission either through LOS signal or through reflections from reflectors. The eavesdroppers can not move their positions. Therefore, success of overhearing BS's transmission heavily depends on the availability of LOS or reflections from the environment. Initially, eavesdropper uses directional antenna and continuously scan
the environment for the best possible reception from BS.The eavesdropper steps through her beam patterns sequentially and decides
on the beam pattern with highest RSSI/SINR from BS to
overhear. The best direction or sector for an eavesdropper could be a LOS or reflection link. The eavesdropper periodically sweeps the environment to
update her best sector to overhear BS to user communication.
The probability of eavesdropper for successfully overhearing
BS's communication depends on the LOS or reflection link
between BS and eavesdropper. Accordingly, in subsequent sections, we derive eavesdropping success probability for an arbitrary Eve by combining ESPs due to the LOS and reflection paths.

Let $SINR_{Eve}$ denote signal-to-noise-plus-interference ratio (SINR) obtained at the eavesdropper. We define $C_{es}: SINR_{Eve}>\beta$ as the event when the SINR of Eve is greater than a threshold $\beta$. Accordingly, the probability that an eavesdropper can overhear the signal from a typical mmWave BS is given by
\begin{equation}
P(C_{es}) = P(SINR_{Eve}>\beta)
\end{equation}

\begin{theorem}
\label{theorem1}
In downlink mmWave SWIPT networks, given a time switching ratio $\eta$ and SINR threshold $\beta$, the eavesdropping success probability (ESP) that an eavesdropper can overhear the signal from a typical mmWave BS under independent eavesdropping attack model is given by (\ref{eq_independent}), as shown at the top of next page. 
\end{theorem}

\newcounter{mytempeqncnt}
\begin{figure*}[!t]
\normalsize
\setcounter{mytempeqncnt}{15}
\setcounter{equation}{15}
\begin{equation}
\label{eq_independent}
\begin{aligned}
P(C_{es}) &= (1-e^{-2\pi \lambda_{n} \frac{e^{-p}}{\beta_{0}^{2}}[1-\beta_{0}a e^{-\beta_{0}a}-e^{-\beta_{0}a}]}) \cdot e^{-k(\beta_{0}d+p)} \\
&\cdot (1-e^{-\lambda_{0}(\mathbb{E}[l]\sqrt{c^{2}\tau^{2}-D^{2}cos^{2}\theta}+\mathbb{E}[w]D|cos\theta|+\mathbb{E}[l]\mathbb{E}[w]-\frac{\mathbb{E}[l](c\tau-D)}{4}-\frac{\mathbb{E}[l]^{2}\sqrt{c^{2}\tau^{2}-D^{2}}}{8D})})\\
&+ (1-e^{-2\pi \lambda_{n} \frac{e^{-p}}{\beta_{0}^{2}}[1-\beta_{0}be^{-\beta_{0}b}-e^{-\beta_{0}b}]}) \cdot (1-e^{-k(\beta_{0}d+p)}) \\
&\cdot e^{-\lambda_{0}(\mathbb{E}[l]\sqrt{c^{2}\tau^{2}-D^{2}cos^{2}\theta}+\mathbb{E}[w]D|cos\theta|+\mathbb{E}[l]\mathbb{E}[w]-\frac{\mathbb{E}[l](c\tau-D)}{4}-\frac{\mathbb{E}[l]^{2}\sqrt{c^{2}\tau^{2}-D^{2}}}{8D})},
\end{aligned}
\end{equation}
\setcounter{equation}{16}
with \\
\begin{equation}
\left\{
\begin{array}{lr}
a=\sqrt[\alpha]{\frac{((1-\eta)P_{t}-\beta \eta P_{t})G_{e}c^2}{(\sigma_{e}^{2}+\sum_{x \in \Phi_{b}/b}I_{x,e}) \beta(4\pi f_{c})^2}} \\
b=\sqrt[\alpha]{\frac{((1-\eta)P_{t}-\beta \eta P_{t})G_{e}c^2}{(\sigma_{e}^{2}+\sum_{x \in \Phi_{b}/b}I_{x,e}) \beta(4\pi f_{c})^2\rho}}
\end{array}
\right.
\end{equation}
\hrulefill
\vspace*{4pt}
\end{figure*}

\begin{proof}
We define two events $LOS_{Eve}$ and $Ref_{Eve}$ as the events
when the eavesdropper is covered by a LOS signal from
the BS and by reflections from the
reflectors present in the environment, respectively. We only consider first-order reflections. By taking into account the possibility
of eavesdropper being covered by either LOS from BS or by
reflections from the environment, the coverage probability of eavesdropper is given by

\begin{equation}
\begin{aligned}
P(C_{sec}) &= P(SINR_{Eve}> \beta|LOS_{Eve})P(LOS_{Eve}) \\
&+P(SINR_{Eve} > \beta)|Ref_{Eve})P(Ref_{Eve})
\end{aligned}
\end{equation}

\subsubsection{Eavesdropping Success Probability due to LOS Link}
We first discuss the success probability of eavesdropping due to LOS link between the typical mmWave BS and an arbitrary eavesdropper, as shown in Figure~\ref{ind_eve_los}.

\begin{figure}[htbp]
	\centering
	\includegraphics[scale=0.33]{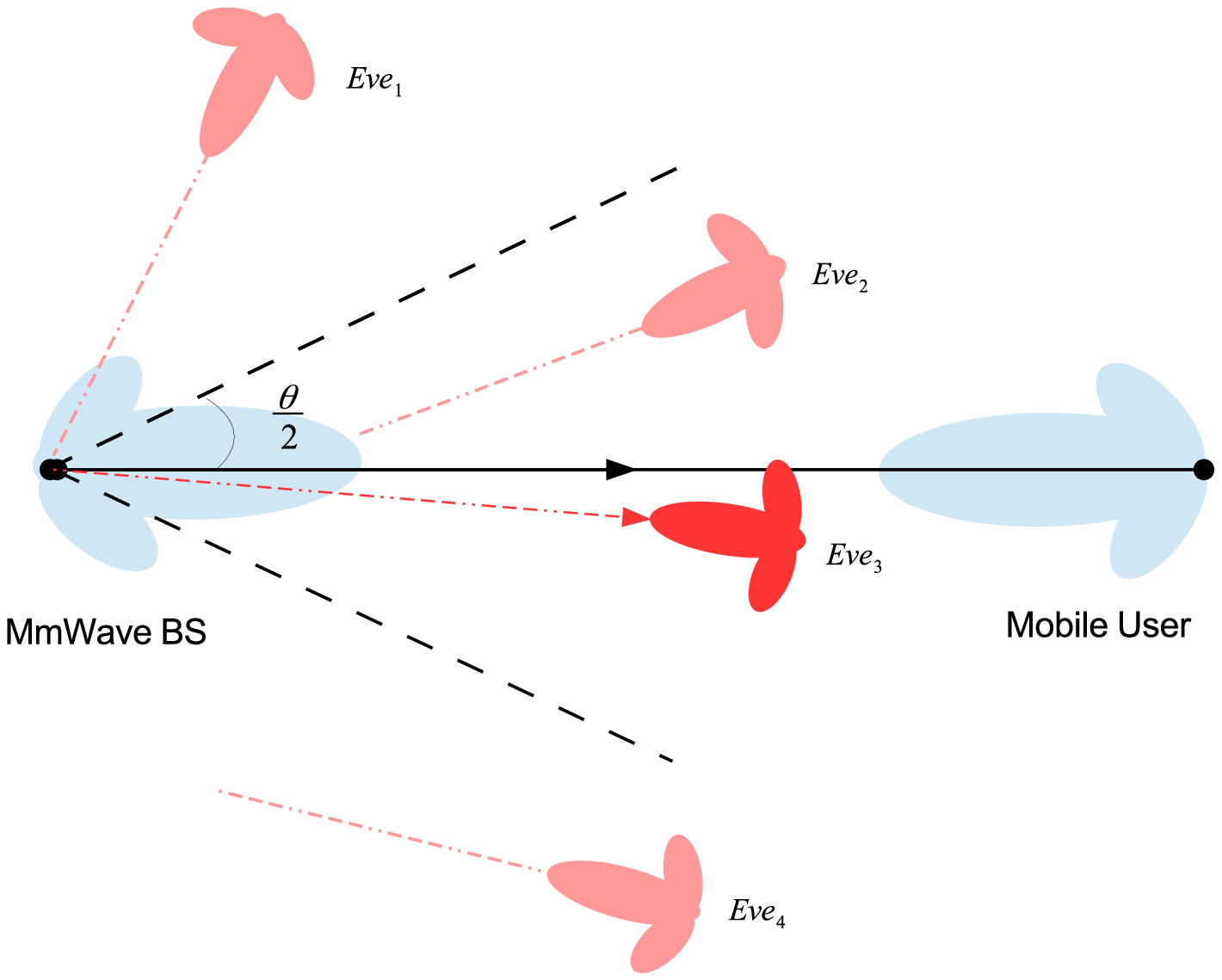}
	\caption{LOS scenario under the independent eavesdropping attack model.}\label{ind_eve_los}
\end{figure}

The signal received at an arbitrary Eve from the targeted mmWave BS is composed of the signal received during time fraction $\eta$, given by
\begin{equation}
y_{e}^{E} = \sum{\bm{h}_{e}^{H}}\sum{\bm{x}^{E}s^{E}}+z_{0},
\end{equation}
and the signal received at the Eve during time fraction $1-\eta$, given by
\begin{equation}
y_{e}^{I} = \sum{\bm{h}_{e}^{H}}\sum{\bm{x}^{I}s^{I}}+z_{0},
\end{equation}
where $\bm{h}_{e}^{H}$ is the wiretap channel matrix between the mmWave BS and Eve, $\bm{x}^{E}$ and $\bm{x}^{I}$ are the EH and ID beamforming vectors used by BS, respective, $z_{0}$ is additive white Gaussian noise.

Since the eavesdropper is not aware of the time switch ratio $\eta$, the signal $y_{e}^{E} $ is considered as an additional noise to jam the eavesdropper.

Moreover, conventionally it is believed that the narrow beam can mitigate the interference between different communication links. However, some millimeter wave experimental measurements have shown that the first order reflections from the environment contribute to majority of signal power in NLOS \cite{Fuschini2017AnalysisOI}. Therefore, in a practical ultra-dense mmWave network, the interference at the Eve from other mmWave BSs can not be neglected. 

Let $S$ denote the power of the incoming signal of interest at Eve, $S$ is given by
\begin{equation}
S = (1-\eta)P_{t} \mathbb{E}[G_{e}]L_{Los}(b,e),
\end{equation}
where $P_{t}$ is the transmission power of mmWave BS, $\mathbb{E}[G_{e}]$ is the expectation of the channel gain that is obtained by Eve, and $L_{Los}(b,e)$ is the path loss function between the mmWave BS and the Eve over LOS path.

Let $I$ denote the interference that is received at Eve, and $I$ is the summation of the interference power from other mmWave BSs' signal and the received power from the targeted BS during time slot $\eta$. Thus, $I$ is given by 
\begin{equation}
\begin{aligned}
I &= I_{b,e}+\sum_{x \in \Phi_{b}/b}I_{x,e} \\
&=\eta P_{t}\mathbb{E}[G_{e}]L_{Los}(b,e)+\sum_{x \in \Phi_{b}/b}P_{t}\mathbb{E}[G_{e}]L(x,e),
\end{aligned}
\end{equation}

Let $N$ denote the number of interference mmWave BSs. Since there exists high path loss in the mmWave signal propagation, the number of $N$ is limited. Let $r$ denote the largest distance between a BS and this Eve, such that Eve can receive signals from this BS. Since the mmWave BSs are modeled as PPP with the density $\lambda_{b}$, the mean number of interference mmWave BSs in the region with center Eve and area $\pi r^{2}$ is given by $\mathbb{E}(N)=\pi r^{2} \lambda_{b}$. Let $x_{i}$ denote the distance between this eavesdropper and the $i$th transmitter in the area $A$, the $x$ is a random variable and the PDF $f_{r}(x)$ of $x$ is given by Lemma \ref{lemma3}.

\begin{figure}[htbp]
	\centering
	\includegraphics[scale=0.33]{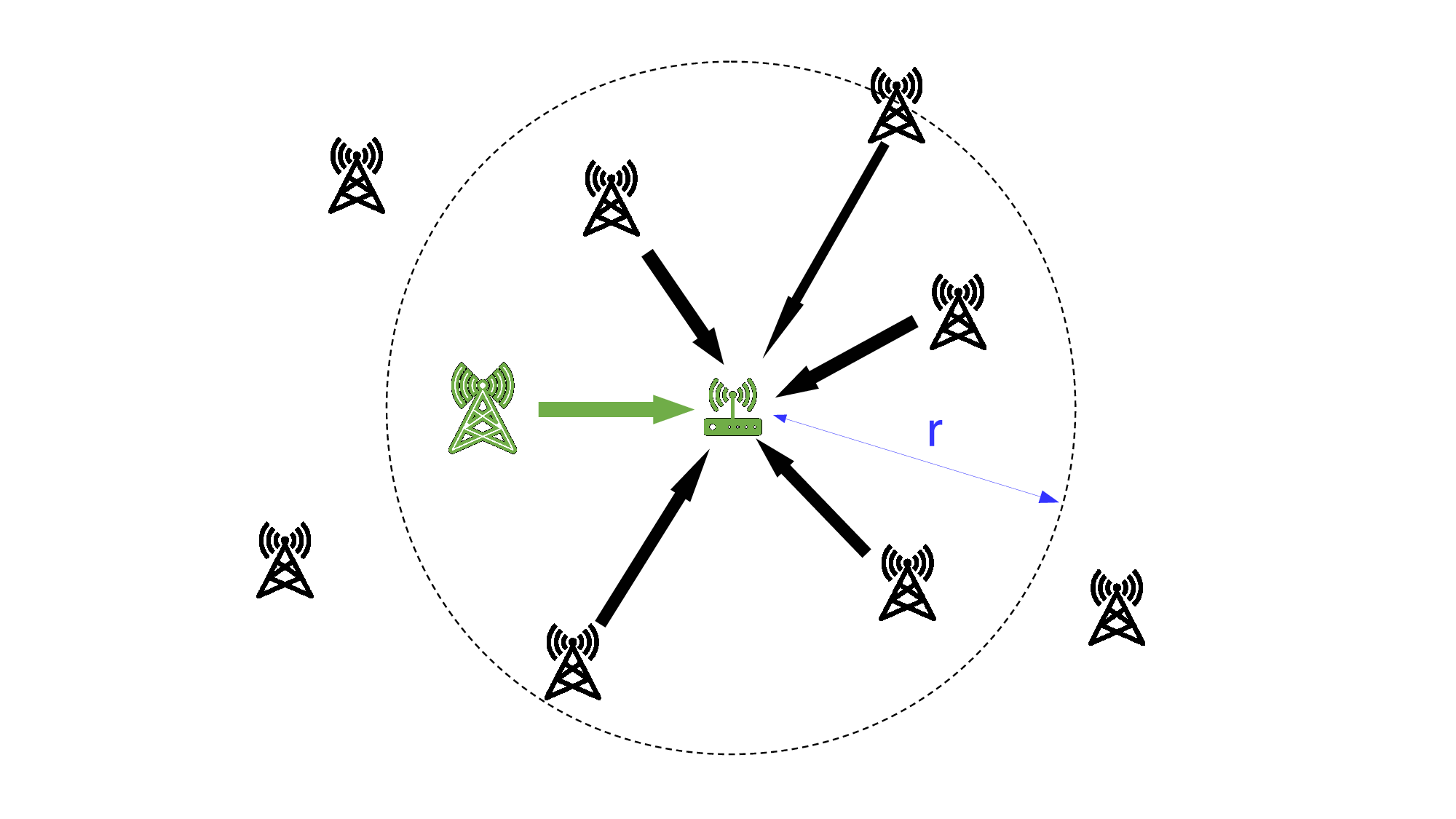}
	\caption{Illustration of interference from other mmWave BSs in area $A$.}\label{interference}
\end{figure}

Therefore, $\sum_{x \in \Phi_{b}/b}I_{x,e}$ can be approximated as
\begin{equation}
\label{appro1}
\begin{aligned}
\sum_{x \in \Phi_{b}/b}I_{x,e} &\approx \pi r^{2} \lambda_{b} \int_{0}^{r}f_{r}(x)P_{t}\mathbb{E}(G_{e})L(x)dx
\end{aligned}
\end{equation}

Note that the interference signal can reach the Eve through LOS and reflection path, so (\ref{appro1}) can be rewritten as
\begin{equation}
\label{appro2}
\begin{aligned}
\sum_{x \in \Phi_{b}/b}I_{x,e} &\approx \pi r^{2} \lambda_{b} P_{t}\mathbb{E}[G_{e}] \int_{0}^{r}(f_{r}(x)(P(LOS)L_{LOS}(x)\\
&+P(Ref)L_{Ref}(x))dx),
\end{aligned}
\end{equation}
where $P(LOS_{Eve})$ and $P(Ref_{Eve})$ are given in (24) and (32), $f_{r}(x)$ is given by Lemma \ref{lemma3}, and $\mathbb{E}(G_{e})$ is given by Corollary \ref{corollary1}, and

\begin{equation}
\label{int_independent}
\left\{
\begin{array}{lr}
\begin{aligned}
\int_{0}^{r}f_{r}(x)L_{LOS}(x)dx &= \int_{0}^{r} \frac{2x}{r^{2}}x^{-\alpha_{L}}dx=\frac{2}{r^{\alpha_{L}}(2-\alpha_{L})}\\
\int_{0}^{r}f_{r}(x)L_{Ref}(x)dx &= \int_{0}^{r} \frac{2x}{r^{2}}x^{-\alpha_{N}}dx=\frac{2}{r^{\alpha_{N}}(2-\alpha_{N})}
\end{aligned}
\end{array}
\right.
\end{equation}

By taking into account the stochastic geometry results in the previous section, the probability that eavesdropper will be in LOS with respect
to BS is given by 
\begin{equation}
P(LOS_{Eve})=P_{Los} \overline{P_{Ref}}
\end{equation}
where $P_{Los}$ and $P_{Ref}$ are given in (\ref{eq_plos}) and (\ref{eq_pref}), respectively.

Therefore, the signal-to-interference-plus-noise ratio (SINR) of the Eve is expressed as 
\begin{equation}
\begin{aligned}
SINR_{Eve} &\triangleq \frac{S}{I+\sigma_{e}^{2}}\\
&=\frac{(1-\eta)P_{t}\mathbb{E}[G_{e}]L(b,e)}{\eta P_{t}\mathbb{E}[G_{e}] L(b,e)+\sum_{x \in \Phi_{b}/b}I_{x,e}+\sigma_{e}^{2}},
\end{aligned}
\end{equation}
where $\sigma_{e}^{2}$ is the noise power at the Eve.

Let $SINR_{Eve}>\beta$, we get 
\begin{equation}
\label{llos}
L(b,e) > \frac{(\sigma_{e}^{2}+\sum_{x \in \Phi_{b}/b}I_{x,e}) \beta}{((1-\eta)P_{t}-\beta \eta P_{t})\mathbb{E}[G_{e}]}
\end{equation}
where $L(b,e)$ is the path loss from the typical BS to the Eve, and 
\begin{equation}
\label{lbe}
L(b,e) = (\frac{c}{4\pi f_{c}})^2 d^{-\alpha}
\end{equation}

From (\ref{llos}) and (\ref{lbe}), we get 
\begin{equation}
d < \sqrt[\alpha]{\frac{((1-\eta)P_{t}-\beta \eta P_{t})\mathbb{E}[G_{e}]c^2}{\sigma_{e}^{2} \beta(4\pi f_{c})^2}}
\end{equation}

Therefore, the secrecy probability due to LOS is given by 
\begin{equation}
\label{eq_p1}
\begin{aligned}
P(SINR_{Eve} & > \beta)|LOS_{Eve})P(LOS_{Eve})\\
&=P(d < \sqrt[\alpha]{\frac{((1-\eta)P_{t}-\beta \eta P_{t})\mathbb{E}[G_{e}]c^2}{(\sigma_{e}^{2}+\sum_{x \in \Phi_{b}/b}I_{x,e}) \beta(4\pi f_{c})^2}}) \\
&\cdot P(LOS_{Eve})\\
&=(F_{d}(x) \mid_{x=a} \cdot P_{Los} \cdot \overline{P_{Ref}} \\
&=(1-e^{-2\pi \lambda_{n} \frac{e^{-p}}{\beta_{0}^{2}}[1-\beta_{0}a e^{-\beta_{0}a}-e^{-\beta_{0}a}]}) \\
&\cdot P_{Los} \cdot \overline{P_{Ref}} \\
\end{aligned}
\end{equation}
where $a=\sqrt[\alpha]{\frac{((1-\eta)P_{t}-\beta \eta P_{t})\mathbb{E}[G_{e}]c^2}{(\sigma_{e}^{2}+\sum_{x \in \Phi_{b}/b}I_{x,e}) \beta(4\pi f_{c})^2}}$, $F_{d}(x)$ is given by Lemma \ref{lemma2}, and $\sum_{x \in \Phi_{b}/b}I_{x,e})$ is given by (\ref{appro2}).

\subsubsection{Secrecy Probability due to Reflection Path}

\begin{figure}[htbp]
	\centering
	\includegraphics[scale=0.33]{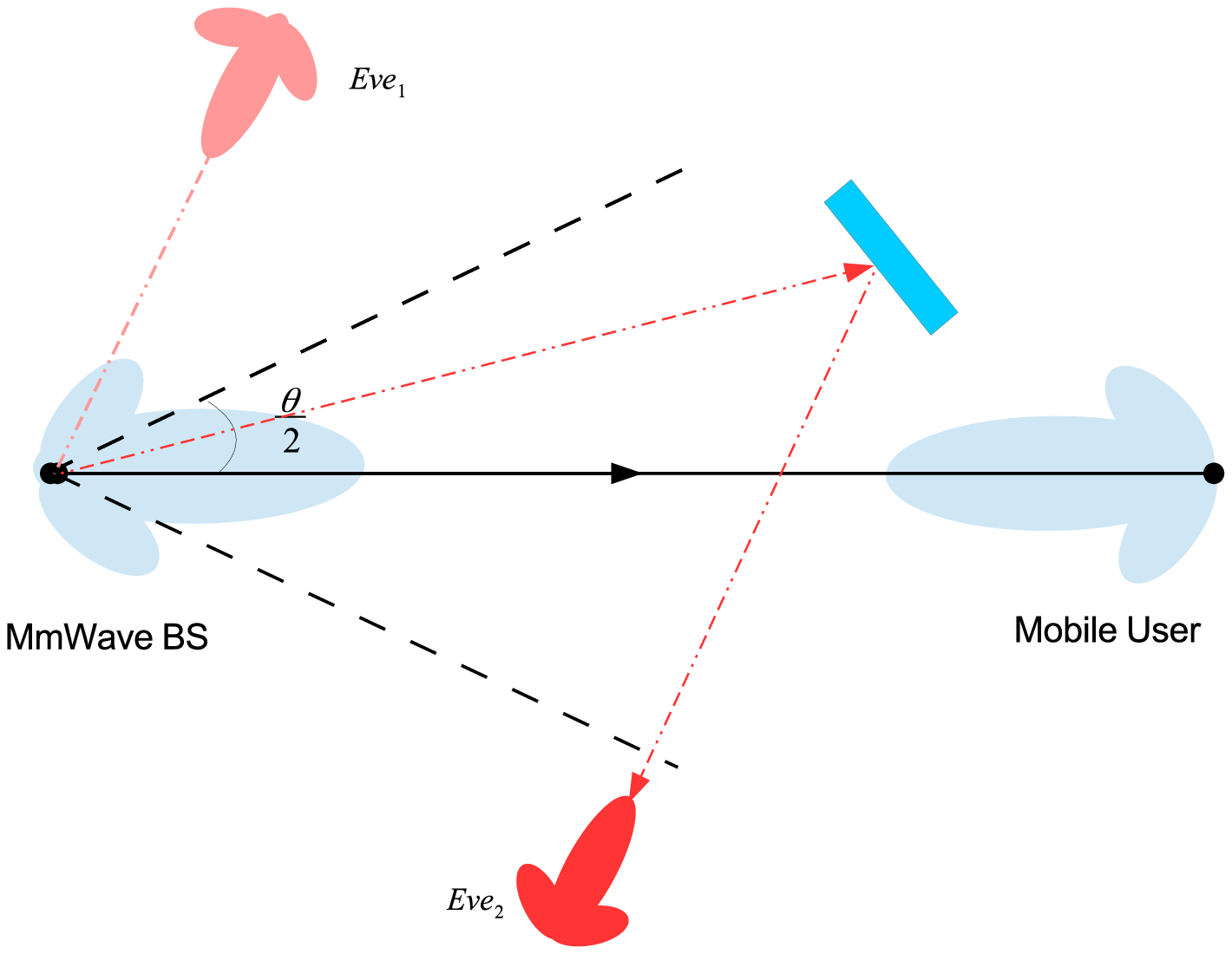}
	\caption{Reflection scenario under the independent eavesdropping attack model.}\label{ind_eve_ref}
\end{figure}

Now, we focus on the derivation of the secrecy probability of BS-User pair due to the reflectors in the environments, as shown in Figure~\ref{ind_eve_ref}.

\begin{equation}
\label{sinr_ref}
\begin{aligned}
SINR_{Eve} &\triangleq \frac{S}{I+\sigma_{e}^{2}}\\
&=\frac{(1-\eta)P_{t} \mathbb{E}[G_{e}]L(b,e)\rho}{\eta P_{t} \mathbb{E}[G_{e}] L(b,e)\rho+\sum_{x \in \Phi_{b}/b}I_{x,e}+\sigma_{e}^{2}}> \beta
\end{aligned}
\end{equation}

From (\ref{sinr_ref}), we obtain 
\begin{equation}
d < \sqrt[\alpha]{\frac{((1-\eta)P_{t}-\beta \eta P_{t})\mathbb{E}[G_{e}]c^2}{(\sum_{x \in \Phi_{b}/b}I_{x,e}+\sigma_{e}^{2}) \beta(4\pi f_{c})^2\rho}}
\end{equation}

By taking into account the stochastic geometry results in the previous section, the probability that eavesdropper will be in a reflection path with respect
to BS is given by 
\begin{equation}
P(Ref_{Eve})=P_{Ref}\overline{P_{Los}}
\end{equation}
where $P_{Los}$ and $P_{Ref}$ are given in (\ref{eq_plos}) and (\ref{eq_pref}), respectively.

Therefore, the secrecy probability due to reflection is given by 
\begin{equation}
\label{eq_p2}
\begin{aligned}
P(SINR_{Eve} & > \beta)|Ref_{Eve})P(Ref_{Eve})\\
&=P(d < \sqrt[\alpha]{\frac{((1-\eta)P_{t}-\beta \eta P_{t})\mathbb{E}[G_{e}]c^2}{(\sum_{x \in \Phi_{b}/b}I_{x,e}+\sigma_{e}^{2}) \beta(4\pi f_{c})^2\rho}})\\
&\cdot P(LOS_{Eve})\\
&=F_{d}(x) \mid_{x=b} \cdot \overline{P_{Los}} \cdot P_{Ref} \\
&=(1-e^{-2\pi \lambda_{n} \frac{e^{-p}}{\beta_{0}^{2}}[1-\beta_{0}be^{-\beta_{0}b}-e^{-\beta_{0}b}]}) \\
&\cdot \overline{P_{Los}} \cdot P_{Ref} \\
\end{aligned}
\end{equation}
where $b=\sqrt[\alpha]{\frac{((1-\eta)P_{t}-\beta \eta P_{t})\mathbb{E}[G_{e}]c^2}{(\sum_{x \in \Phi_{b}/b}I_{x,e}+\sigma_{e}^{2}) \beta(4\pi f_{c})^2\rho}}$.

The total secrecy probability due to LOS and reflection links is the summation of (\ref{eq_p1}) and (\ref{eq_p2}), which is given by (\ref{eq_independent}).

\end{proof}

\subsection{Colluding Eavesdroppers}

\begin{figure}[htbp]
	\centering
	\includegraphics[scale=0.33]{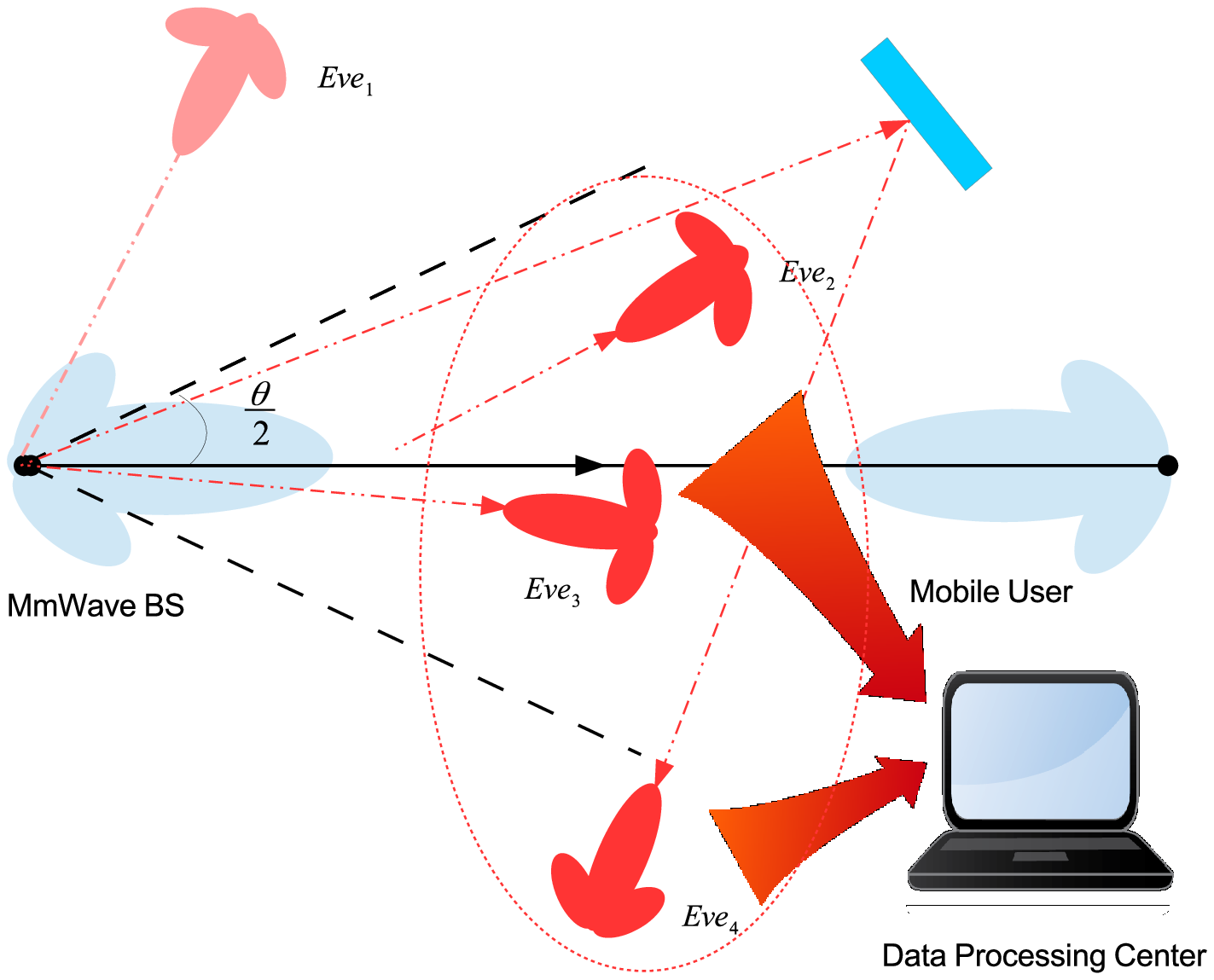}
	\caption{Colluding eavesdropping attack model.}\label{colluding_eve}
\end{figure}

In this part, we will consider the colluding eavesdroppers, which can jointly process their received signals at a central data processing unit, as shown in Figure~\ref{colluding_eve}, . In this model, we consider the eavesdroppers adopt the maximal-ratio combining (MRC) scheme to get the most confidential information and the best possible performance. Therefore, the eavesdropping success probability for the colluding eavesdroppers is given by
\begin{equation}
P(C_{es}) = P(SINR_{Eves\_Col}> \beta)
\end{equation}

\newcounter{mytempeqncnt2}
\begin{figure*}[!t]
\normalsize
\setcounter{mytempeqncnt2}{36}
\setcounter{equation}{36}
\begin{equation}
\label{eq_colluding}
P(C_{es}) \approx P(\frac{(1-\eta)(P_{Los} \overline{P_{Ref}} \cdot \frac{2}{r^{\alpha_{L}}(2-\alpha_{L})}+\rho \overline{P_{Los}} P_{Ref} \cdot \frac{2}{r^{\alpha_{N}}(2-\alpha_{N})}}{(\pi r^{2}\lambda_{b}+\eta)(P_{Los} \overline{P_{Ref}}  \cdot \frac{2}{r^{\alpha_{L}}(2-\alpha_{L})}+ \rho \overline{P_{Los}} P_{Ref} \cdot \frac{2}{r^{\alpha_{N}}(2-\alpha_{N})}}>\beta)
\end{equation}
\setcounter{equation}{37}
with \\
\begin{equation}
\left\{
\begin{array}{lr}
P_{Los}=e^{-k(\beta_{0}d+p)} \\
P_{Ref} =exp(\lambda_{0}(\mathbb{E}[l]\sqrt{c^{2}\tau^{2}-D^{2}cos^{2}\theta}+\mathbb{E}[w]D|cos\theta|+\mathbb{E}[l]\mathbb{E}[w]-\frac{\mathbb{E}[l](c\tau-D)}{4}-\frac{\mathbb{E}[l]^{2}\sqrt{c^{2}\tau^{2}-D^{2}}}{8D})) 
\end{array}
\right.
\end{equation}
\hrulefill
\vspace*{4pt}
\end{figure*}

\begin{theorem}
\label{theorem2}
In downlink mmWave SWIPT networks, given a time switching ratio $\eta$ and SINR threshold $\beta$, the secrecy probability of a typical mmWave base station and mobile user pair under colluding eavesdropping attack model is given by (\ref{eq_colluding}), as shown at the top of this page.
\end{theorem}

\begin{proof}
The total SINR that is obtained by all the eavesdroppers is given by
\begin{equation}
\label{sinr_colluding_1}
\begin{aligned}
SINR_{Eves\_Col} &= \frac{S}{I+\sigma_{e}^{2}}\\
&=\frac{S_{Los}+S_{Ref}}{I_{Los}+I_{Los}+I_{x,e}+\sigma_{e}^{2}}
\end{aligned}
\end{equation}
with
\begin{equation}
\label{sinr_colluding_2}
\left\{
\begin{array}{lr}
S_{Los} = \sum (1-\eta)Pt \mathbb{E}[G_{e}]L(b,e), \\
S_{Ref} = \sum (1-\eta)Pt \mathbb{E}[G_{e}]L(b,e)\rho, \\
I_{Los} = \sum \eta Pt \mathbb{E}[G_{e}]L(b,e), \\
I_{Ref} = \sum \eta Pt \mathbb{E}[G_{e}]L(b,e)\rho,  \\
I_{x,e} = \sum_{e \in \Phi_{e}}\sum_{x \in \Phi_{b}/b} Pt \mathbb{E}[G_{e}]L(x, e).
\end{array}
\right.
\end{equation}

As the derivation in the previous subsection, (\ref{sinr_colluding_2}) can be approximated as 
\begin{equation}
\label{sinr_colluding_3}
\left\{
\begin{array}{lr}
\begin{aligned}
S_{Los} &\approx (1-\eta)\pi r^{2}\lambda_{e}P_{t} \mathbb{E}[G_{e}] P_{Los} \overline{P_{Ref}} \\
&\int_{0}^{r} f_{r}(x)L_{Los}(x)dx,\\
S_{Ref} &\approx (1-\eta)\pi r^{2}\lambda_{e}P_{t} \mathbb{E}[G_{e}] \overline{P_{Los}} P_{Ref}\rho \\
&\int_{0}^{r} f_{r}(x)L_{Ref}(x)dx, \\
I_{Los} &\approx \eta \pi r^{2}\lambda_{e}P_{t} \mathbb{E}[G_{e}] P_{Los} \overline{P_{Ref}} \int_{0}^{r} f_{r}(x)L_{Los}(x)dx,\\
I_{Ref} &\approx \eta \pi r^{2}\lambda_{e}P_{t} \mathbb{E}[G_{e}] \overline{P_{Los}} P_{Ref}\rho \int_{0}^{r} f_{r}(x)L_{Ref}(x)dx, 
\end{aligned}
\end{array}
\right.
\end{equation}
And
\begin{equation}
\label{sinr_colluding_4}
\begin{aligned}
I_{x,e} &\approx \pi r^{2} \lambda_{e} \pi r^{2} \lambda_{b} P_{t}\mathbb{E}[G_{e}] \int_{0}^{r}(f_{r}(x)(x)(P(LOS)L_{LOS}(x)\\
&+P(Ref)L_{Ref}(x))dx),
\end{aligned}
\end{equation}
where $ \mathbb{E}[G_{e}]$ is given by Corollary \ref{corollary1}.

Compared with the received power from all the mmWave BSs, the white noise in the environment can be neglected. Substituting (\ref{sinr_colluding_3}) and (\ref{sinr_colluding_4}) into (\ref{sinr_colluding_1}), we obtain

\begin{equation}
\label{sinr_colluding_5}
\begin{aligned}
SINR_{Eves_Col} &\approx (1-\eta)(P_{Los} \overline{P_{Ref}} \int_{0}^{r} f_{r}(x)L_{Los}(x)dx \\
&+\overline{P_{Los}} P_{Ref}\rho \int_{0}^{r} f_{r}(x)L_{Ref}(x)dx)\\
&/((\pi r^{2}\lambda_{b}+\eta)(P_{Los} \overline{P_{Ref}} \int_{0}^{r} f_{r}(x)L_{Los}(x)dx \\
&+ \overline{P_{Los}} P_{Ref}\rho \int_{0}^{r} f_{r}(x)L_{Ref}(x)dx))
\end{aligned}
\end{equation}

Let $SINR_{Eves_Col}>\beta$, we obtain
\begin{equation}
\label{sinr_colluding_final}
\begin{aligned}
&(1-\eta)(P_{Los} \overline{P_{Ref}} \cdot \frac{2}{r^{\alpha_{L}}(2-\alpha_{L})} +\rho \overline{P_{Los}} P_{Ref}  \cdot \frac{2}{r^{\alpha_{N}}(2-\alpha_{N})}\\
&/((\pi r^{2}\lambda_{b}+\eta)(P_{Los} \overline{P_{Ref}}  \cdot \frac{2}{r^{\alpha_{L}}(2-\alpha_{L})} + \rho \overline{P_{Los}} P_{Ref} \cdot \frac{2}{r^{\alpha_{N}}(2-\alpha_{N})}) > \beta
\end{aligned}
\end{equation}

Then, the equation (\ref{eq_colluding}) is derived.
\end{proof}
\section{Simulation}
In this section, we present our numerical results and simulation validation methodology of our proposed eavesdropping attack in Section \ref{mm_esp}.

\label{mm_results}
\subsection{Simulation Environment and Setup}

To validate the analytical findings in this work, we develop simulation programs based on an open-source mmWave channel simulator named NYUSIM, which is developed by researchers at New York University (NYU) \cite{7996792}\cite{ju2019millimeterwave}. NYUSIM is developed based on extensive real-world wideband propagation channel measurements at multiple mmWave frequencies from 28 GHz to 73 GHz. Those channel measurements are widely obtained from various outdoor environments in urban microcell, urban macrocell, and rural macrocell environments. This simulator can generate realistic temporal and spatial channel responses to support realistic physical layer simulations and design for 5G communications. The latest 2.0 version of NUYSIM that is published in August, 2019 implements the simulation of human blockage shadowing loss the reflection in the signal propagation. The base code is written in MATLAB, and the users can create their own codes based on this open-source simulator.

\begin{figure}[htbp]
	\centering
	\includegraphics[scale=0.4]{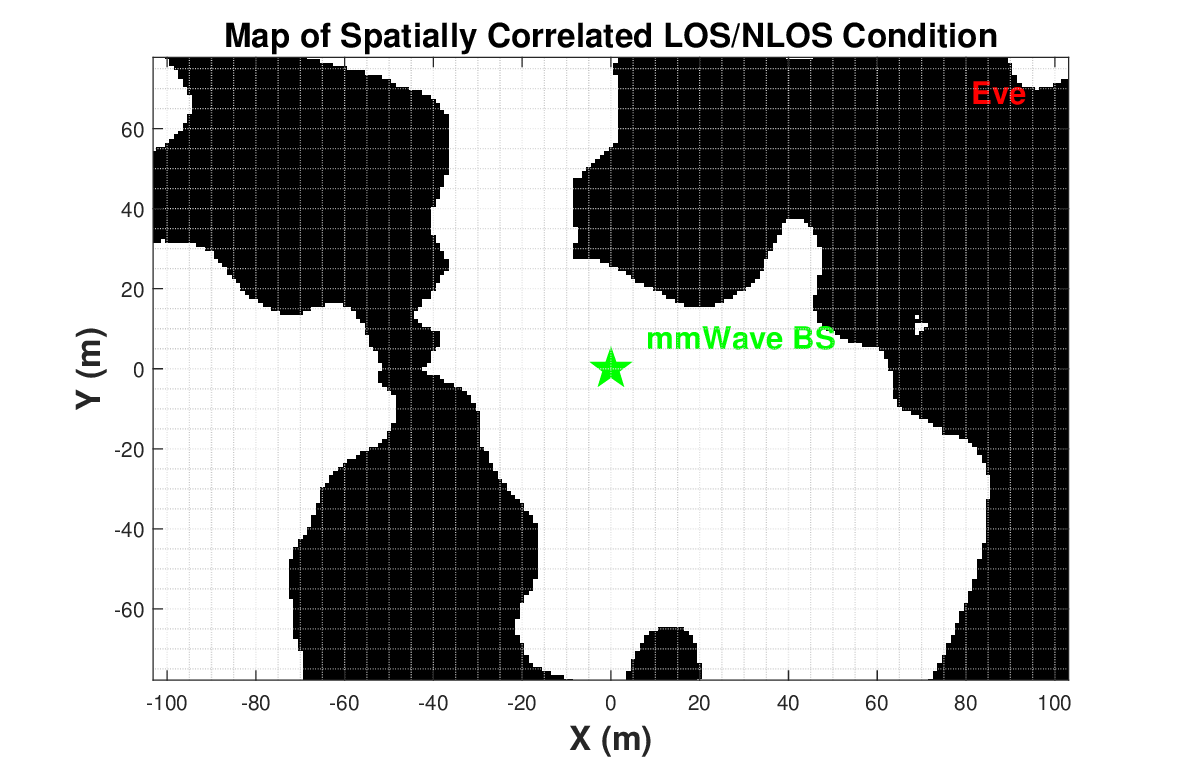}
	\caption{A randomly generated 2-D $200m \times 160m$ map for LOS/Reflection condition in UMi scenario. The granularity of the map is $1m$. The height of the mmWave BS and eavesdropper are $35m$ and $1.5m$, respectively. The signal may propagate along the LOS or reflection path according to the eavesdropper's location in the map.}\label{sim_map1}
\end{figure}

In the simulation, we randomly generate 2-dimensional (2D) grid maps in a simulated area, where the obstacles and reflectors are randomly distributed. One of the maps is shown in Figure \ref{sim_map1}, where the "white" represents LOS whereas the "black" represents reflection. Since we consider the urban micro-cell (UMi) networks in this work, the simulated area is set to be $200m \times 160m$. The granularity of the map is $1m$. The height of the mmWave BS and eavesdropper are $35m$ and $1.5m$, respectively. The signal may propagate along the LOS or reflection path according to the eavesdropper's location in the map. For each map, we perform 1000 simulation runs (i.e., set the number of eavesdroppers' locations to 1000) to emulate 1000 random mmWave channel realizations with the input parameters described in Table \ref{tab:2}. For each simulation run, the location of the eavesdropper is random, and the distance between the mmWave BS and the eavesdropper varies from the lower bound and upper bound as shown in Table \ref{tab:2}.

\begin{table}[h]
\caption{Important Parameters}
\centering
\begin{tabular}{l | l}
\hline
Distance Range Option & 10-500m \\
RF bandwidth & 800 MHz \\
Scenario & UMi \\
Lower Bound of T-R Separation Distance & 10m \\
Upper Bound of T-R Separation Distance & 500m \\
Tx Power & 30dBm \\
mmWave Bs Height & 35m \\
Eavesdropper Height & 1.5m \\
Temperature & $20^{\circ}C$ \\
Humidity & $50\%$ \\
Rain Rate & 0 mm/hr \\
Number of RX locations & 1000 \\
TX Array Type & ULA \\
RX Array Type & ULA \\
Number of TX Antenna Elements & N=2 \\
Number of RX Antenna Elements & N=2 \\
Main-lobe gain & N \\
Side-lobe gain & $\frac{1}{sin^2(3\pi/2\sqrt{N})}$ \\
\hline
\end{tabular}
\label{tab:2}
\end{table}

In our numerical and simulation analysis, we use different mmWave carrier frequencies on 28GHz, 38GHz, and 60GHz, in which their LOS and NLOS path loss exponents are shown in Table \ref{tab:1} based on the practical channel measurements, and noise power is $\sigma_{w}^{2}=-174+10\log_{10}10+10dB$. The ULA is used at all nodes, and the beamwidth $\theta$, main-lobe gain, and side-lobe gain of an $N$ elements ULA are shown in Table \ref{tab:1}.

\begin{table}[h]
\caption{Path Loss Exponent for mmWave Channels \cite{7880674}\cite{7247348}\cite{6363891}}
\centering
\begin{tabular}{l | l | l| l}
Path Loss exponent & 28GHz & 38GHz & 60GHz \\
\hline
LOS & 2 & 2 & 2.25\\
Reflection & 3 & 3.71 & 3.76 \\
\hline
\end{tabular}
\label{tab:1}
\end{table}

Based on the results from the $1000 \times 2$ random simulation runs, the SINRs of eavesdroppers calculated from the received power, and the probabilities of these SINRs are greater than a pre-defined threshold $\beta$ are obtained. And $\beta$ varies from $1dB$ to $100dB$. The numerical and simulation results are shown in the following subsection.

\subsection{Numerical and Simulation Results}
In this subsection, some numerical and simulation results are provided to validate the accuracy of the theoretical analysis and illustrate the impact of different system parameters on the eavesdropping success probability of eavesdroppers. 

\begin{figure}[htbp]
	\centering
	\includegraphics[scale=0.6]{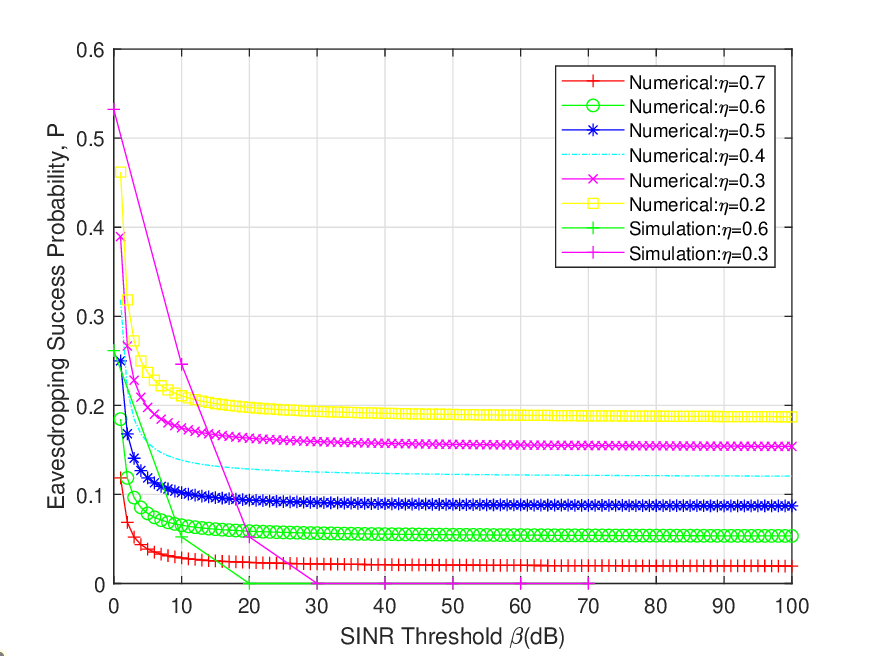}
	\caption{Secrecy probability versus SINR threshold $\beta$ for different time switching ratio $\eta$ under independent eavesdropping. $N_{b}=3$, $f_{c}=28 GHz$.}\label{result_ind_eta}
\end{figure}

Figure~\ref{result_ind_eta} plots the eavesdropping success probability (ESP) of an arbitrary eavesdropper versus the SINR threshold $\beta$ for different time switching ratio $\eta$ under the independent eavesdropping. We first observe that the ESP of eavesdropper significantly decreases with the increase in the threshold $\beta$, and varies indistinctively with the increasing of $\beta$ when $\beta > 10 dB$. Second, for a fixed SINR threshold $\beta$, the ESP can be improved when the time switching ratio $\eta$ decreases. This is because when $\eta$ decreases, the BS would spend less energy to transmit the power, and the signal eavesdroppers receive for energy harvest is the additional interference to eavesdroppers. Therefore, the decrease of time switch ratio increases the SINR that is obtained by eavesdroppers. The insight from this result is that if the legitimate mmWave BSs and mobile users want to achieve more secure communication, the BS needs to increase the ratio $\eta$ and spend more energy on the power transfer. However, the increase of $\eta$ will lead to the decrease of data transfer. Therefore, how to balance the secrecy performance and data transfer rate will be a research direction in the future.

Figure~\ref{result_ind_eta} also compares the numerical and simulation results. Note that the values of ESPs in numerical  and  simulation  results are not identical. ESPs  in  simulation results  are  greater  than  the  ones  in  numerical  results under the same parameters. The reason is that we adopt the ideal sector antenna model in our theoretical analysis; however, in the simulation, the antennas are not ideal sectored to capture the practical beam patterns. Therefore, eavesdroppers are able to receive stronger signals from the sidelobe of beam. However, the ESPs in both numerical and simulation results show similar trends in this figure. In fact, the theoretical analysis using ideal antenna model is also valuable, which provides the conservative lower bounds of ESPs for eavesdroppers when the threshold $\beta$ is small. The success probabilities of eavesdropping can not be less than these bounds in practice. On the other hand, in the perspectives of legitimate BSs and users, the theoretical analysis tells them the upper bounds of their secrecy probabilities.

\begin{figure}[htbp]
	\centering
	\includegraphics[scale=0.6]{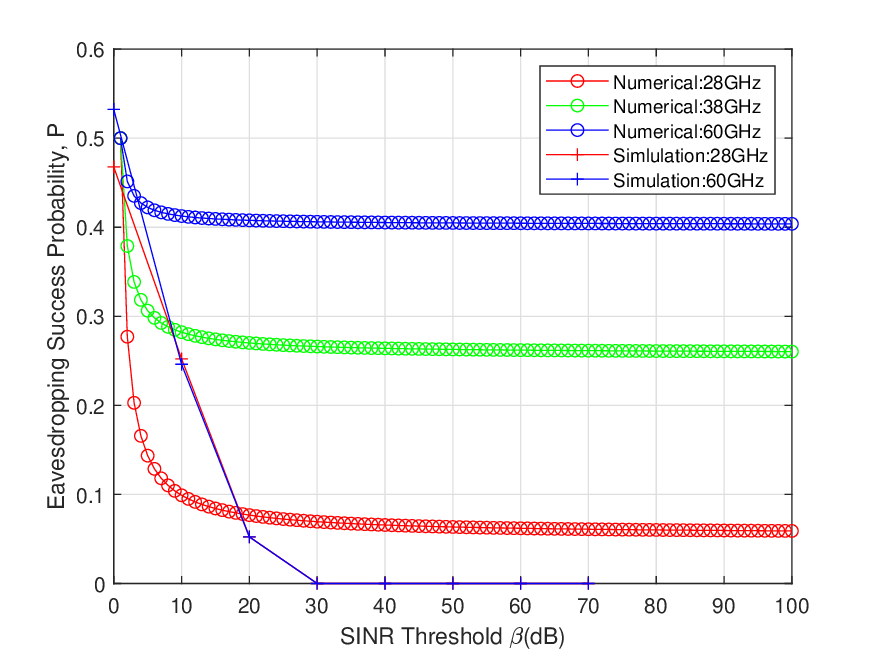}
	\caption{Secrecy probability versus SINR threshold $\beta$ for different carrier's frequency under the independent eavesdropping. $N_{b}=3$, $\eta=0.3$.}\label{result_ind_fc}
\end{figure}

Figure~\ref{result_ind_fc} presents the ESP versus SINR threshold $\beta$ for different carrier's frequencies under the independent eavesdropping. From this figure, our numerical and simulation results show that the ESP increases with the increasing of $f_{c}$ in the independent eavesdropper model. As shown in this figure, for a fixed SINR threshold $\beta$, the ESP can be improved when the mmWave carrier frequency increases. Note that there exists significant differences of ESPs for different carrier's frequencies in the numerical results, i.e., when $\beta=8dB$ and $f_{c}=60GHz$, $ESP=0.42$, and when $\beta=8dB$ and $f_{c}=28GHz$, $ESP=0.11$. On the other hand, differences of ESPs for different carrier's frequencies are indistinctive in the simulation results. That's because for simplicity, we use the uniform channel model for different carrier's frequencies in our analytical derivation. However, the channels are more complicated in practice for different mmWave frequencies. Unfortunately, to date, there is no any standard mmWave channel model, which is also the primary challenge in the physical layer design in the next generation 5G networks \cite{7414384}.

\begin{figure}[htbp]
	\centering
	\includegraphics[scale=0.6]{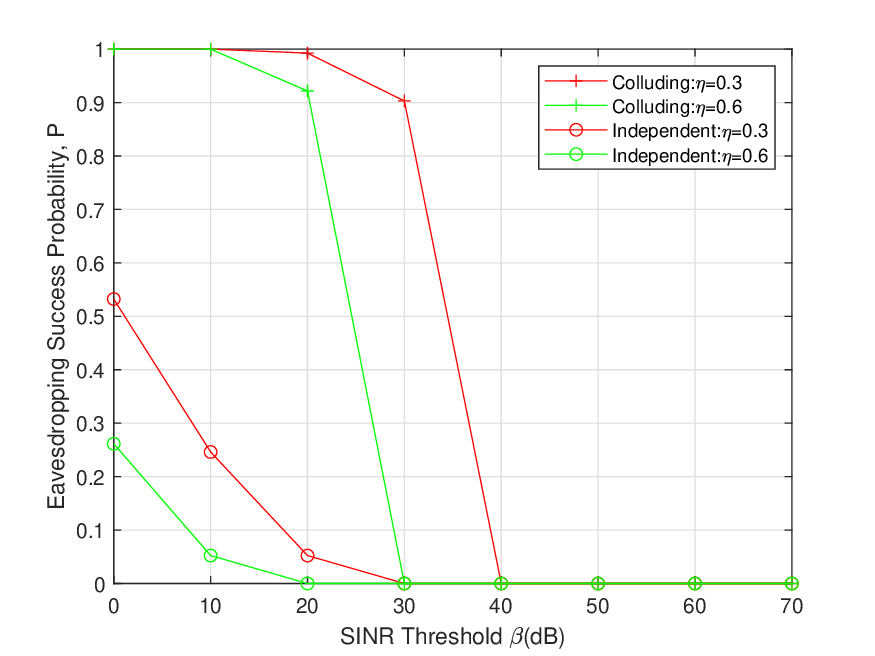}
	\caption{Secrecy probability versus SINR threshold $\beta$ for different time switching ratio $\eta$ under the colluding eavesdropping. $N_{b}=3$, $f_{c}=28 GHz$.}\label{result_col_eta}
\end{figure}

From the Theorem \ref{theorem2}, we get the insight that only the radius $r$, the density of mmWave BSs $\lambda_{b}$, and time switching ratio $\eta$ affect the ESP in the colluding eavesdropping scenario. And the radius $r$, the density of mmWave BSs $\lambda_{b}$ are relevant to the number of BSs $N_{b}$ (target BS plus interference BSs). Thus, in the following, we now investigate effects of time switching ratio $\eta$ and the number of interference BSs $N$ on the ESP of eavesdroppers, under the colluding eavesdropping strategy. In Figure \ref{result_col_eta}, the eavesdropping success probability versus SINR threshold $\beta$ for different time switching ratio $\eta$ under the colluding eavesdropping is plotted. For comparison, we also plot the ESP for the same time switching ratios under the independent eavesdropping strategy. From this figure, We can see that the ESP of eavesdropper decreases with the increasing of the threshold $\beta$. And for a fixed SINR threshold $\beta$, the ESP can be improved when the time switching ratio decreases. Moreover, it is illustrated that under the colluding eavesdropping strategy, eavesdroppers have significant higher probability of overhearing BS's transmission than under the independent eavesdropping strategy.

\begin{figure}[htbp]
	\centering
	\includegraphics[scale=0.57]{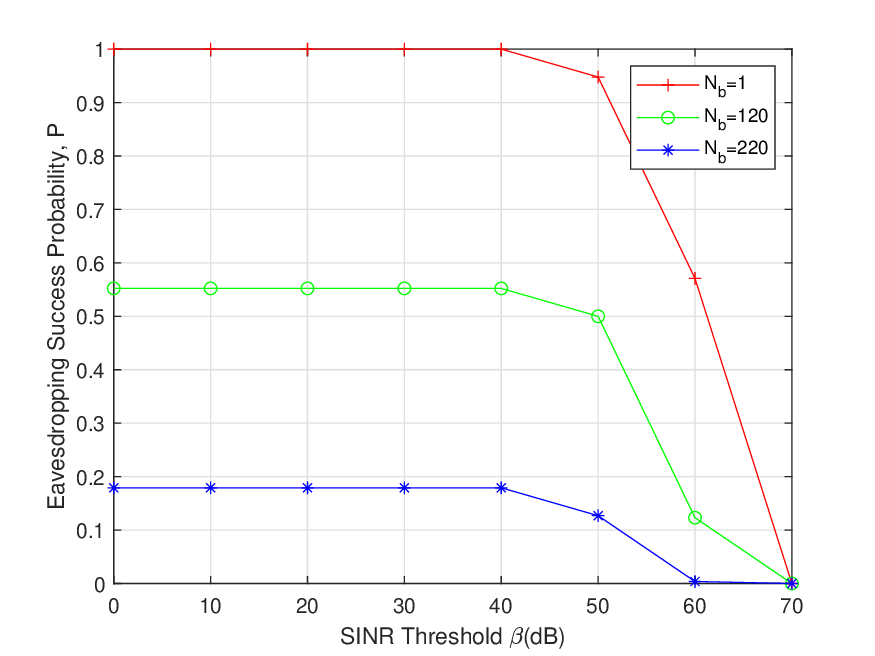}
	\caption{Secrecy probability versus SINR threshold $\beta$ for different number of mmWave BSs under the colluding eavesdropping. $\eta=0.4$, $f_{c}=28 GHz$.}\label{result_col_n}
\end{figure}

Figure \ref{result_col_n} shows the effect of the number of mmWave BSs $N_{b}$ (target BS plus interference BSs) in the communication range of eavesdroppers on the ESP. For a fixed SINR threshold $\beta$, the ESP increases with the number $N_{b}$ decreases. As an extreme case, when $N_{b}=1$, i.e., there is no interference from other mmWave BSs, the colluding eavesdroppers have $100\%$ probability to overhear the transmission from the target BS, even when SINR threshold $\beta=40dB$. The insight from this observation is that we should deploy smaller cell (i.e., more mmWave BSs in a certain area) in practice to achieve more secure communication in the mmWave SWIPT networks.

\section{Conclusion}
\label{mm_conclusion}
In this paper, we have studied the security performance in general mmWave SWIPT networks. Analytical expressions of eavesdropping success probability (ESP) of both independent and colluding eavesdroppers have been derived, by incorporating the random reflection paths in the environment. Our analysis has modeled two types of eavesdropping attack strategies to the mmWave SWIPT networks, namely independent eavesdropping and colluding eavesdropping attacks. We have derived the analytical expressions for eavesdropping success probability (ESP) of eavesdroppers under independent and colluding eavesdropping attacks, considering both the LOS and reflection paths of signal propagation. To get insights, we have statistically examined the security performance trends in terms of key parameters such as the time switching strategy in SWIPT, densities of mmWave BSs, carriers frequencies and time switching ratio at BSs, etc. Our numerical and simulation results have revealed the effects of some key parameters on the security performance. Based on the numerical and simulation results, some design strategies of mmWave SWIPT (e.g., selections of transmission parameters) have been suggested to defend against eavesdropping attacks and achieve secure communication in practice.


\bibliographystyle{IEEEtran}
\bibliography{citations}

\begin{thebibliography}{10}
\providecommand{\url}[1]{#1}
\csname url@samestyle\endcsname
\providecommand{\newblock}{\relax}
\providecommand{\bibinfo}[2]{#2}
\providecommand{\BIBentrySTDinterwordspacing}{\spaceskip=0pt\relax}
\providecommand{\BIBentryALTinterwordstretchfactor}{4}
\providecommand{\BIBentryALTinterwordspacing}{\spaceskip=\fontdimen2\font plus
\BIBentryALTinterwordstretchfactor\fontdimen3\font minus
  \fontdimen4\font\relax}
\providecommand{\BIBforeignlanguage}[2]{{%
\expandafter\ifx\csname l@#1\endcsname\relax
\typeout{** WARNING: IEEEtran.bst: No hyphenation pattern has been}%
\typeout{** loaded for the language `#1'. Using the pattern for}%
\typeout{** the default language instead.}%
\else
\language=\csname l@#1\endcsname
\fi
#2}}
\providecommand{\BIBdecl}{\relax}
\BIBdecl

\bibitem{6740844}
A.~{Zanella}, N.~{Bui}, A.~{Castellani}, L.~{Vangelista}, and M.~{Zorzi},
  ``Internet of things for smart cities,'' \emph{IEEE Internet of Things
  Journal}, vol.~1, no.~1, pp. 22--32, Feb 2014.

\bibitem{8214104}
T.~D. {Ponnimbaduge Perera}, D.~N.~K. {Jayakody}, S.~K. {Sharma},
  S.~{Chatzinotas}, and J.~{Li}, ``Simultaneous wireless information and power
  transfer (swipt): Recent advances and future challenges,'' \emph{IEEE
  Communications Surveys Tutorials}, vol.~20, no.~1, pp. 264--302, Firstquarter
  2018.

\bibitem{7346844}
D.~{Steinmetzer}, J.~{Chen}, J.~{Classen}, E.~{Knightly}, and M.~{Hollick},
  ``Eavesdropping with periscopes: Experimental security analysis of highly
  directional millimeter waves,'' in \emph{2015 IEEE Conference on
  Communications and Network Security (CNS)}, Sep. 2015, pp. 335--343.

\bibitem{7862142}
Y.~{Ju}, H.~{Wang}, T.~{Zheng}, and Q.~{Yin}, ``Secure transmissions in
  millimeter wave systems,'' \emph{IEEE Transactions on Communications},
  vol.~65, no.~5, pp. 2114--2127, May 2017.

\bibitem{7880674}
Y.~{Zhu}, L.~{Wang}, K.~{Wong}, and R.~W. {Heath}, ``Secure communications in
  millimeter wave ad hoc networks,'' \emph{IEEE Transactions on Wireless
  Communications}, vol.~16, no.~5, pp. 3205--3217, May 2017.

\bibitem{7876781}
M.~E. {Eltayeb}, J.~{Choi}, T.~Y. {Al-Naffouri}, and R.~W. {Heath}, ``Enhancing
  secrecy with multiantenna transmission in millimeter wave vehicular
  communication systems,'' \emph{IEEE Transactions on Vehicular Technology},
  vol.~66, no.~9, pp. 8139--8151, Sep. 2017.

\bibitem{7464352}
C.~{Wang} and H.~{Wang}, ``Physical layer security in millimeter wave cellular
  networks,'' \emph{IEEE Transactions on Wireless Communications}, vol.~15,
  no.~8, pp. 5569--5585, Aug 2016.

\bibitem{7505974}
S.~{Vuppala}, S.~{Biswas}, and T.~{Ratnarajah}, ``An analysis on secure
  communication in millimeter/micro-wave hybrid networks,'' \emph{IEEE
  Transactions on Communications}, vol.~64, no.~8, pp. 3507--3519, Aug 2016.

\bibitem{7178504}
C.~{Rusu}, N.~{González-Prelcic}, and R.~W. {Heath}, ``An attack on antenna
  subset modulation for millimeter wave communication,'' in \emph{2015 IEEE
  International Conference on Acoustics, Speech and Signal Processing
  (ICASSP)}, April 2015, pp. 2914--2918.

\bibitem{8723482}
S.~{Balakrishnan}, P.~{Wang}, A.~{Bhuyan}, and Z.~{Sun}, ``Modeling and
  analysis of eavesdropping attack in 802.11ad mmwave wireless networks,''
  \emph{IEEE Access}, vol.~7, pp. 70\,355--70\,370, 2019.

\bibitem{8387202}
L.~{Wang}, K.~{Wong}, S.~{Jin}, G.~{Zheng}, and R.~W. {Heath}, ``A new look at
  physical layer security, caching, and wireless energy harvesting for
  heterogeneous ultra-dense networks,'' \emph{IEEE Communications Magazine},
  vol.~56, no.~6, pp. 49--55, June 2018.

\bibitem{7491259}
T.~A. {Khan}, A.~{Alkhateeb}, and R.~W. {Heath}, ``Millimeter wave energy
  harvesting,'' \emph{IEEE Transactions on Wireless Communications}, vol.~15,
  no.~9, pp. 6048--6062, Sep. 2016.

\bibitem{Wang_2017}
\BIBentryALTinterwordspacing
L.~Wang, K.-K. Wong, R.~W. Heath, and J.~Yuan, ``Wireless powered dense
  cellular networks: How many small cells do we need?'' \emph{IEEE Journal on
  Selected Areas in Communications}, vol.~35, no.~9, p. 2010–2024, Sep 2017.
  [Online]. Available: \url{http://dx.doi.org/10.1109/jsac.2017.2720858}
\BIBentrySTDinterwordspacing

\bibitem{7997004}
L.~{Wang} and K.~{Wong}, ``Energy coverage in wireless powered sub-6 ghz and
  millimeter wave dense cellular networks,'' in \emph{2017 IEEE International
  Conference on Communications (ICC)}, May 2017, pp. 1--6.

\bibitem{8500123}
R.~{Zhu}, H.~{Fu}, and T.~{Shu}, ``Information-theoretic security and energy
  efficiency for information and power transfer in two-hop wireless relay
  networks,'' in \emph{2018 IEEE International Conference on
  Electro/Information Technology (EIT)}, 2018, pp. 0877--0881.

\bibitem{8566013}
X.~{Sun}, W.~{Yang}, Y.~{Cai}, L.~{Tao}, Y.~{Liu}, and Y.~{Huang}, ``Secure
  transmissions in wireless information and power transfer millimeter-wave
  ultra-dense networks,'' \emph{IEEE Transactions on Information Forensics and
  Security}, vol.~14, no.~7, pp. 1817--1829, July 2019.

\bibitem{6932503}
T.~{Bai} and R.~W. {Heath}, ``Coverage and rate analysis for millimeter-wave
  cellular networks,'' \emph{IEEE Transactions on Wireless Communications},
  vol.~14, no.~2, pp. 1100--1114, Feb 2015.

\bibitem{5783993}
Z.~{Pi} and F.~{Khan}, ``An introduction to millimeter-wave mobile broadband
  systems,'' \emph{IEEE Communications Magazine}, vol.~49, no.~6, pp. 101--107,
  June 2011.

\bibitem{1580787}
F.~{Baccelli}, B.~{Blaszczyszyn}, and P.~{Muhlethaler}, ``An aloha protocol for
  multihop mobile wireless networks,'' \emph{IEEE Transactions on Information
  Theory}, vol.~52, no.~2, pp. 421--436, Feb 2006.

\bibitem{article}
G.~Yang, J.~Du, and M.~Xiao, ``Analysis on 60 ghz wireless communications with
  beamwidth-dependent misalignment,'' 11 2016.

\bibitem{6835179}
J.~{Wildman}, P.~H.~J. {Nardelli}, M.~{Latva-aho}, and S.~{Weber}, ``On the
  joint impact of beamwidth and orientation error on throughput in directional
  wireless poisson networks,'' \emph{IEEE Transactions on Wireless
  Communications}, vol.~13, no.~12, pp. 7072--7085, Dec 2014.

\bibitem{6840343}
T.~{Bai}, R.~{Vaze}, and R.~W. {Heath}, ``Analysis of blockage effects on urban
  cellular networks,'' \emph{IEEE Transactions on Wireless Communications},
  vol.~13, no.~9, pp. 5070--5083, Sep. 2014.

\bibitem{Muhammad2017AnalyticalMF}
N.~A. Muhammad, P.~Wang, Y.~Li, and B.~Vucetic, ``Analytical model for outdoor
  millimeter wave channels using geometry-based stochastic approach,''
  \emph{IEEE Transactions on Vehicular Technology}, vol.~66, pp. 912--926,
  2017.

\bibitem{Fuschini2017AnalysisOI}
F.~Fuschini, S.~Haefner, M.~Zoli, R.~M{\"u}ller, E.~M. Vitucci, D.~Dupleich,
  M.~Barbiroli, J.~Luo, E.~Schulz, V.~Degli-Esposti, and R.~Thomae, ``Analysis
  of in-room mm-wave propagation: Directional channel measurements and ray
  tracing simulations,'' \emph{Journal of Infrared, Millimeter, and Terahertz
  Waves}, vol.~38, pp. 727--744, 2017.

\bibitem{7996792}
S.~{Sun}, G.~R. {MacCartney}, and T.~S. {Rappaport}, ``A novel millimeter-wave
  channel simulator and applications for 5g wireless communications,'' in
  \emph{2017 IEEE International Conference on Communications (ICC)}, May 2017,
  pp. 1--7.

\bibitem{ju2019millimeterwave}
S.~Ju, O.~Kanhere, Y.~Xing, and T.~S. Rappaport, ``A millimeter-wave channel
  simulator nyusim with spatial consistency and human blockage,'' 2019.

\bibitem{7247348}
S.~{Deng}, M.~K. {Samimi}, and T.~S. {Rappaport}, ``28 ghz and 73 ghz
  millimeter-wave indoor propagation measurements and path loss models,'' in
  \emph{2015 IEEE International Conference on Communication Workshop (ICCW)},
  June 2015, pp. 1244--1250.

\bibitem{6363891}
T.~S. {Rappaport}, E.~{Ben-Dor}, J.~N. {Murdock}, and Y.~{Qiao}, ``38 ghz and
  60 ghz angle-dependent propagation for cellular peer-to-peer wireless
  communications,'' in \emph{2012 IEEE International Conference on
  Communications (ICC)}, June 2012, pp. 4568--4573.

\bibitem{7414384}
M.~{Agiwal}, A.~{Roy}, and N.~{Saxena}, ``Next generation 5g wireless networks:
  A comprehensive survey,'' \emph{IEEE Communications Surveys Tutorials},
  vol.~18, no.~3, pp. 1617--1655, thirdquarter 2016.

\end{thebibliography}

\end{document}